\documentclass[fleqn,usenatbib]{mnras}

\usepackage{newtxtext,newtxmath}

\usepackage[T1]{fontenc}
\usepackage{ae,aecompl}

\usepackage{graphicx}	

\newcommand{\wno}{\,cm$^{-1}$\,}	

\title[Infrared spectra of chain silicates]{Infrared Spectra of Pyroxenes (Crystalline Chain Silicates) at Room Temperature}

\author[Bowey, Hofmeister \& Keppel]{J. E. Bowey,$^{1}$\thanks{E-mail:boweyj@cardiff.ac.uk}
A. M. Hofmeister$^{2}$ \& E. Keppel$^{2}$ \\
$^{1}$ School of Physics and Astronomy, Cardiff University, Queen's Buildings, The Parade, Cardiff, Wales,
CF24 3AA\\
$^{2}$Department of Earth and Planetary Sciences, Washington University, 1
Brookings Drive, St. Louis MO 63130, USA.\\
}

\date{Accepted XXX. Received YYY; in original form ZZZ}

\pubyear{2020}

\begin{document}
\label{firstpage}
\pagerange{\pageref{firstpage}--\pageref{lastpage}}
\maketitle

\begin{abstract}
Crystals of pyroxene are common in meteorites but few compositions
have been recognized in astronomical environments due to the limited
chemistries included in laboratory studies. We present quantitative
room-temperature spectra of 17 Mg-- Fe-- and Ca--bearing ortho- and
clinopyroxenes, and a Ca-pyroxenoid in order to discern trends
indicative of crystal structure and a wide range of composition. Data
are produced using a Diamond Anvil Cell: our band strengths are up to
6 times higher than those measured in KBr or polyethylene dispersions,
which include variations in path length (from grain size) and surface
reflections that are not addressed in data processing. Pyroxenes have
varied spectra: only two bands, at 10.22~$\mu$m and 15.34~$\mu$m in
enstatite (En$_{99}$), are common to all. Peak-wavelengths generally
increase as Mg is replaced by Ca or Fe. However, two bands in
MgFe-pyroxenes shift to shorter wavelengths as the Fe component
increases from 0 to 60 per cent. A high-intensity band shifts from
11.6~$\mu$m to 11.2~$\mu$m and remains at 11.2~$\mu$m as Fe increases
to 100~per~cent; it resembles an astronomical feature normally
identified with olivine or forsterite. The distinctive pyroxene bands
between 13~ and 16~$\mu$m show promise for their identification in
\emph{Mid-Infrared-Instrument (MIRI)} spectra obtained with the
\emph{James Webb Space Telescope (JWST)}. The many pyroxene bands
between 40 and 80~$\mu$m could be diagnositic of silicate mineralogy
if data were obtained with the proposed \emph{Space Infrared Telescope
  for Cosmology and Astrophysics (SPICA)}. Our data indicate that
comparison between room-temperature laboratory bands for enstatite and
cold $\sim 10-K$ astronomical dust features at wavelengths $\gtrsim
28~\mu$m can result in the identification of (Mg,Fe)- pyroxenes that
contain 7--15~per~cent less Fe-- than their true values because some
temperature shifts mimic some compositional shifts. Therefore some
astronomical silicates may contain more Fe, and less Mg, than previously
thought.
\end{abstract}




\section{Introduction}

Silicate dust is ubiquitous in galaxies; before the mid-1990s
amorphous, or glassy, submicron-sized silicate grains were assigned as
carriers of broad strong absorption and emission bands centred at 10
and 18~$\mu$m in dusty environments including the envelopes of evolved
stars, the interstellar medium and comets. Later small fractions of
crystalline silicates were tentatively identified in oxygen-rich
circumstellar environments \citep{Waters1996} and firmly identified in
comets \citep{Wooden1999} and young stellar objects (YSOs)
\citep{Malfait1998} in far-IR spectra obtained by ESA's \emph{Infrared
  Space Observatory} Satellite.  More recently, \emph{Spitzer Infrared
  Spectrograph (IRS)} spectra have revealed crystalline silicate
absorption bands in ultra-luminous IR galaxies \citep{Spoon2006} and
OH-megamasing galaxies \citep{Willett2011}. A systematic search of
7.5--38~$\mu$m \emph{Spitzer} spectra for crystalline silicates finds
that 868 spectra of 790 sources including early-type stars, evolved
stars and galaxies contain crystalline silicates \citep{Chen2016};
spectra show evidence of olivine (forsterite), pyroxene (enstatite) or
other unidentified crystalline minerals. However, many of these
identifications are tentative because relatively few minerals have
been compared with astronomical spectra despite the wide variety of
crystalline silicates found in meteorites and on Earth \citep[see
  e.g.][for a fuller discussion]{BA2002}. Astronomical grains are
mostly expected to be magnesium-rich; whilst this is likely,
laboratory data for magnesium-poor chemical compositions are limited
to one or two compositions \cite[e.g.][]{TMB2009} and the spectra not
readily available, so alternatives have not been much explored by
astrophysicists. \citeauthor{BA2002} took a different approach, and
matched relatively smooth astronomical 10$\mu$m profiles with a
mixture of crystalline silicates with a relatively small component of
amorphous silicate.  By this analysis, the enhanced breadth of the
10-$\mu$m features in the circumstellar discs surrounding young stars
and in molecular clouds could be reproduced by a mixture of
crystalline pyroxenes with varying stoichiometries (80 per cent by
mass) and amorphous silicates (20 per cent) but the laboratory data
and ground-based data were of insufficiently high quality to
rigorously test this hypothesis. Recently, \citet{Do-Duy2020} have
proposed that the enhanced breadth of the 10~$\mu$m feature in
molecular clouds and young stellar objects is due the presence of a
few per cent of crystalline silicate dust (forsterite). The potential
of mixtures of other crystalline silicate groups contributing has not
been considered recently, even though the meteoritic record indicates
a wide variety of crystalline silicates are present in at least some
of these environments. Pyroxenes are a likely candidate because they
are amongst the most common silicates in the most primitive
meteorites.

\subsection{Laboratory studies}

Transmission spectra of pyroxenes embedded in KBr or polyethylene
pellets have been previously published by \citet{Ferraro1982},
\citet{Jaeger1998} and \citet{Chihara2002} but laboratory studies in
the astronomical literature contain a relatively limited range of
compositions including magnesium and iron: 4 Mg-rich Mg--Fe-pyroxenes
\citep{Jaeger1998}, 7 Mg-rich and 1 Fe-pyroxene
\citep{Chihara2002}. Micro-spectroscopic and chemical
techniques have been used to obtain 8--17-$\mu$m spectra of individual
meteoritic grains, covering the Mg-rich end of the series
\citep{BMG2007}. We present new spectra of 8 Mg- and Fe-bearing
orthopyroxenes, 9 Mg-, Ca- and Fe-bearing clinopyroxenes, and one
Ca-bearing pyroxenoid obtained at room temperature over the wavelength
range of the fundamental infrared bands for the range of chemistries
extant in these solid-solution series.

Data were collected from powdered samples compressed in a diamond
anvil cell (DAC) in order to form thin films of uniform thickness with
minimimal spaces between the grains (no embedding medium is
used). Compressing the powder reduces the effect of scattering because
the reflection at grain boundaries is minimized. Grain-sizes are
deduced experimentally, to give the sharpest and most repeatable
infrared bands \citep{HKS2003}. Excess pressure is released prior to
measuring the spectra so that the cell acts only as a sample-holder
(see \citeauthor{HKS2003} and references therein for the detailed
methodology). The method produces nearly perfect absorption spectra
similar to those derived from the reflectivity of single crystals,
with only slight rounding of the strongest peaks.

Our wavelength-absorption coefficients can be used as an optical depth
profile and directly compared with astronomical observations of
optically-thin emission, and foreground absorption in the infrared in
order to derive the chemical and temperature properties of the
dust\footnote{An alternative method is to use absorption coefficients
  from indices of refraction, however, infrared spectra of strong
  spectral features calculated from refractive indices are often
  wavelength-shifted in relation to those of the laboratory spectra of
  particulates \citep[e.g.][]{Jaeger1998, BH1983,
    Sogawa2006}. Problems with particulate data have been resolved, or
  shown to be due to misunderstandings of mineralogy and problems with
  the laboratory technique \citep{SHB1999, HKS2003, Imai2009,
    BAS2000}.}; once these factors are better constrained reflectivity
studies of a narrower range of species can be used to deduce grain
sizes and shapes from weak spectral peaks.  Room-temperature DAC
spectra are available for olivines \citep{Pitman2010}, hydrosilicates
\citep{HB2006}, silicate overtones \citep{BH2005} and glasses
\citep{SWH2011}.

To encourage consistent nomenclature between the disciplines of
mineralogy, astrophysics and chemistry, we describe the mineralogical
nomenclature of pyroxenes in Section~\ref{sec:pyroxenes}; the samples,
experimental and data processing details are in Section~\ref{sec:expt}
where we also discuss the Diamond Anvil Cell (DAC) technique and present a
method for estimating absorption coefficients from thin films of
compressed powder. The complete spectra are presented in frequency
units in Section~\ref{sec:over}, and detailed compositional-wavelength
shifts are discussed and quantified in Section~\ref{sec:shifts}. The
new data are compared with published room-temperature and
low-temperature particulate spectra in Section~\ref{sec:compare}. The
data and initial astronomical implications are summarised in
Section~\ref{sec:summary}.


\section{Pyroxenes}
\label{sec:pyroxenes}
\begin{figure}
\includegraphics*[bb=45 595 270 750]{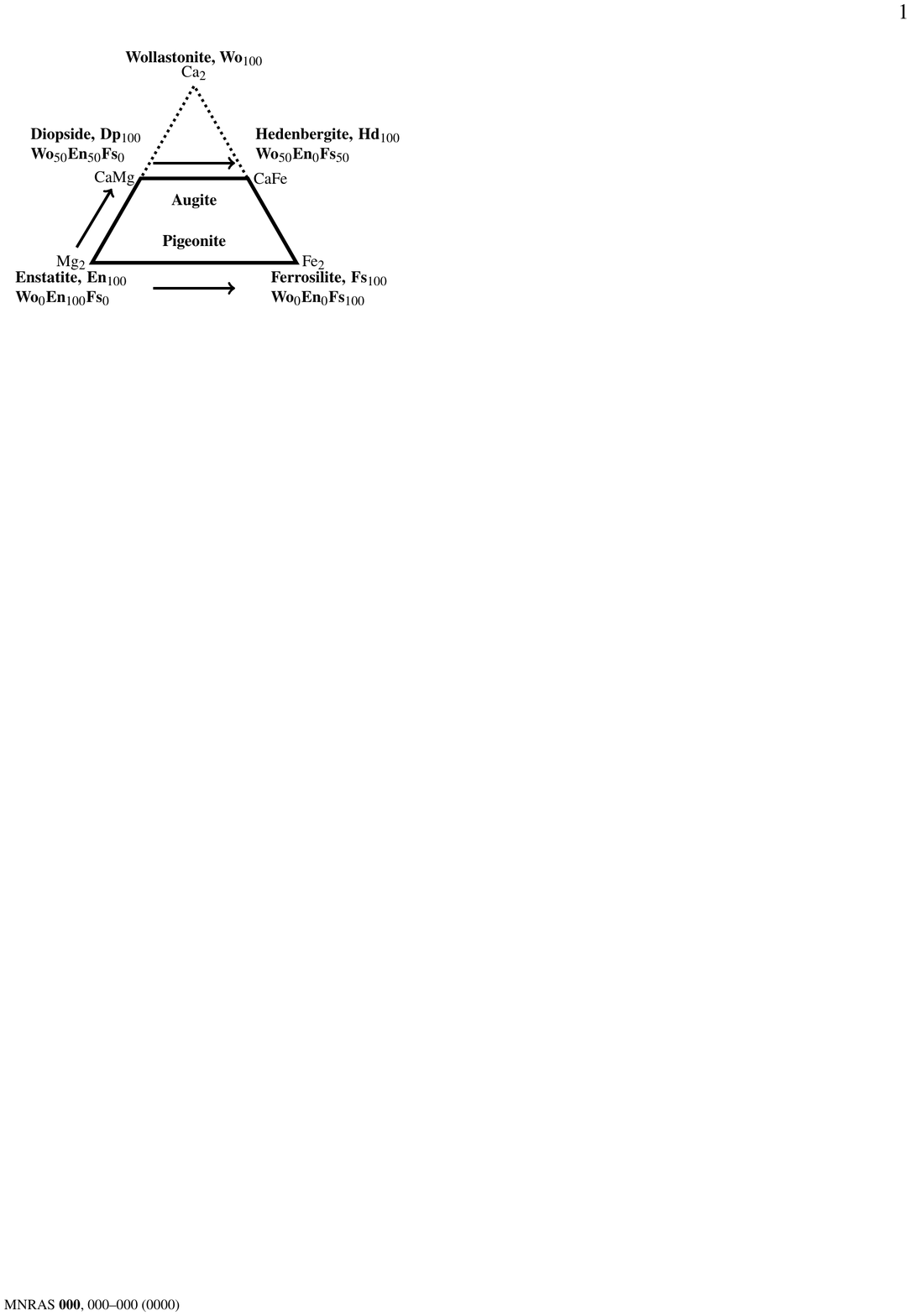}
\caption{Nomenclature and metal components of pyroxenes, chain
  silicates of formula [Metal]$_2$--Si$_2$O$_6$: solid-solution series
  studied in this paper are Enstatite--Ferrosilite (En--Fs
  orthopyroxenes), and monoclinic pyroxenes in the Enstatite--Diopside
  (En--Dp) and Diopside--Hedenbergite (Dp--Hd) series; pigeonite and
  augite have intermediate Mg/Fe ratios with increasing Ca
  content. Wollastonite is a pyroxenoid. \label{fig:nomenclature}}
\end{figure}
\begin{table*}
\begin{minipage}{\linewidth}
  \caption{Sample descriptions of the 18 minerals studied.  For the
    natural samples, the chemical formulae are computed from electron
    microprobe analyses obtained from similar samples at the same
    locality, the formulation of the synthetics is described in the references.  Mass density, $\rho$, is required for
    the derivation of mass extinction coefficients (see
    Section~\ref{sec:massext}, equation~\ref{eq:tau}) from the
    measured natural log absorption coefficients described in
    Section~\ref{sec:bandstr}, equation~\ref{eq:mabs}.\label{tab:samples}}
\begin{tabular}{llclll}
\hline
Sample & Chemical Formula&Ref. &Locality& (Source\footnote{NJ~=~N. Johnson; NM~=~National Museum of Natural History sample \#R18682;ExM~=~Excalibur Mineral Co.; AT~=~Alan Turnock; WU~=~Washington University; SM~=~Schooler's Minerals; RFD~=~R. F. Dymek sample \#72816; SG~=~S. Guggenheim; GRR~=~G. R. Rossman})&$\rho$~gcm$^{-3}$\\ \hline

\multicolumn{5}{l}{\bf Enstatite--Ferrosilite Series (Orthopyroxenes): Fe$^{2+}$ substitution for Mg$^{2+}$}\\
En$_{99}$&(Mg$_{1.98}$Ca$_{0.02}$Fe$_{0.01}$)Si$_{2}$O$_{6}$&\footnote{See \citet{Hofmeister2012} for microprobe analyses: The aubrite contains 5\% microscopic diopside inclusions (blebs) of composition Mg$_ {1.10}$Ca$_{0.85}$Na$_ {0.02}$Al$_{0.02}$)Si$_2$O$_6 \equiv$ Wo$_{43}$En$_{55}$. The Mogok sample is a large, gem-quality single-crystal. Near-IR spectra of the O-H stretching bands and visible to ultraviolet spectra of the electronic transitions (e.g. Fe2+) are also provided of these materials.\label{foot:hof2012}}\footnote{\citet{Okada:1988}}& Norton Co. Aubrite meteorite &(SM) &3.21\\ 

En$_{92}$ &(Mg$_{1.84}$Fe$_{0.11}$Cr$_{0.01}$Ni$_{0.01}$Al$_{0.01}$Ca$_{0.01}$)(Al$_{0.03}$Si$_{1.97}$)O$_{6}$& \ref{foot:hof2012} \footnote{\citet{DHZ1978}; clinopyroxene and pigeonite mass densities, $\rho$ are from
  tabulated values and are quoted to 3 significant figures if they are
  from the same locality, or to two significant figures if they are
  estimates based on a similar composition.\label{foot:DHZ}} & Webster, Jackson, NC &(WU)&3.26\\
En$_{90}$&(Mg$_{1.80}$Fe$_{0.15}$Ca$_{0.02}$Al$_{0.02}$)(Al$_{0.01}$Si$_{1.99}$)O$_{6}$&\ref{foot:hof2012}&Mogok, Burma &(Gemwow)&3.28\\
En$_{85}$&   (Mg$_{1.70}$Fe$_{0.28}$Ca$_{0.01}$)Si$_{2}$O$_{6}$& \footnote{\citet{GR1979}\label{foot:PCM}}\ref{foot:hof2012} & Bamble\footnote{This is the sample incorrectly called \emph{'Bramble'} Enstatite in \citet{Bowey2001}.}, Norway &(Wards) &3.32\\
En$_{55}$&   (Mg$_{1.10}$Fe$_{0.79}$Ca$_{0.07}$Mn$_{0.02}$Al$_{0.01}$Ti$_{0.01}$)Si$_{2.0}$O$_{6}$&{\ref{foot:PCM}}& Summit Rock, OR &(Wards)&3.56\\

En$_{40}$&   (Mg$_{0.80}$Fe$_{1.09}$Mn$_{0.1}$Ca$_{0.03}$)Si$_2$O$_{6}$&\footnote{Microprobe analysis performed at Washington University; analysis in Table~\ref{tab:analysis}.\label{foot:WU}} & Wilagedera, Ceylon &(SG)&3.68 \\ 
En$_{12}$& (Mg$_{0.24}$Fe$_{1.7}$Ca$_{0.04}$Al$_{0.02}$)(Al$_{0.02}$Si$_{1.98}$)O$_{6}$&\ref{foot:PCM}& Greenland &(GRR)&3.90\\ 
En$_1$& (Mg$_{0.02}$Fe$_{1.79}$Mn$_{0.13}$Ca$_{0.03}$Na$_{0.01}$)Si$_{2}$O$_{6}$&\footnote{Sample \#433 \citep{Ormaasen1977}}&Lofoten, Norway &(GRR)&3.99\\
\hline


\multicolumn{5}{l}{\bf Enstatite--Diopside--Wollastonite Series (Clinopyroxenes): Ca$^{2+}$ substitution for Mg$^{2+}$}\\
Wo$_{21}$En$_{79}$&Dp$_{42}$En$_{58}$&\footnote{Dp$_{59}$En$_{41}$ by decomposition of natural tremolite \citep{Johnson2003}, others with similar methodology but different starting compositions \citep{Johnson2002} and personal communication. \label{foot:NJ}}&Synthetic &(NJ)&3.3$^{\ref{foot:DHZ}}$ \\ 
Wo$_{30}$En$_{70}$&Dp$_{59}$En$_{41}$&\ref{foot:NJ}&Synthetic &(NJ)&3.3 \\ 
Wo$_{40}$En$_{60}$&Dp$_{81}$En$_{19}$&\ref{foot:NJ}&Synthetic &(NJ)&3.3\\ 
Wo$_{50}$En$_{50}$&Dp$_{100}$&\ref{foot:NJ}&Synthetic &(NJ)&3.3 \\ 
Wo$_{99}$En$_1$& Ca$_{1.98}$(Mg$_{0.02}$Fe$_{0.01}^{3+}$)(Si$_{1.99}$Al$_{0.01}$)O$_{6}$&\ref{foot:DHZ} &Crestmore, CA &(WU)&2.92 \\ 

\hline

\multicolumn{5}{l}{\bf Diopside-- Hedenbergite Series (Clinopyroxenes): Fe$^{2+}$ substitution for Mg$^{2+}$ with 50~per~cent Ca$^{2+}$}\\
Wo$_{50}$En$_{47}$& (Ca$_{0.99}$Na$_{0.03}$)(Mg$_{0.94}$Fe$_{0.03}$Al$_{0.01}$)Si$_2$O$_{6}$&\footnote{\citet{Hemingway1998}, \citet{HP2008}}&DeKalb, NY &(NM)&3.3$^{\ref{foot:DHZ}}$ \\ 
Wo$_{49}$En$_{36}$& (Ca$_{0.99}$Na$_{0.01}$)(Mg$_{0.73}$Mn$_{0.02}$Fe$_{0.25}$Ti$_{0.01}$Al$_{0.01}$)(Si$_{1.95}$Al$_{0.05}$)O$_{6}$&\ref{foot:WU}&Calumet, CO &(WU)&3.4 \\

Wo$_{47}$En$_6$& Ca$_{0.93}$(Mg$_{0.12}$Mn$_{0.04}$Fe$_{0.80}^{2+}$Fe$_{0.06}^{3+}$Al$_{0.02}$)(Si$_{1.99}$Al$_{0.01}$)O$_{6}$&\ref{foot:WU}& Iona Is., Rockland, NY &(ExM)&3.5\\ 
\hline
\multicolumn{4}{l}{\bf Pigeonite}\\

Wo$_{10}$En$_{62}$& Ca$_{0.20}$(Mg$_{1.25}$Fe$_{0.54}$Al$_{0.02}$)(Si$_{1.98}$Al$_{0.02}$)O$_{6}$&\footnote{A. Turnock, personal communication; \citet{HT1980}}&Synthetic &(AT)&3.3  \\ 
Wo$_{36}$En$_{37}$&(Ca$_{0.72}$Na$_{0.02}$)(Mg$_{0.75}$Mn$_{0.01}$Fe$_{0.48}$Ti$_{0.03}$Al$_{0.01}$)(Si$_{1.93}$Al$_{0.07}$)O$_{6}$&\ref{foot:WU}&Belmont Quarry, Louden, VA &(ExM)&3.4$^{\ref{foot:DHZ}}$\\ 
\hline
\end{tabular}
\end{minipage}
\end{table*}

Pyroxenes are chain silicates, characterized by chains of SiO$_4$
tetrahedra which are linked along the crystallographic c-axis by
shared oxygen atoms. The generic formula is (M2 , M1)T$_2$O$_6$ where
M2 refers to cations in a distorted octahedral coordination, M1 to
cations in a regular octahedral coordination and T to tetrahedrally
coordinated cations (usually Si$^{4+}$) \citep{Morimoto1988}. The
triangle in Figure~\ref{fig:nomenclature} illustrates the nomenclature
of chain silicates in the pyroxene and pyroxenoid groups, with the
most common substitutions. Minerals in the quadrilateral at the bottom
of the figure are denoted pyroxenes. In pyroxenes Mg$^{2+}$, Fe$^{2+}$
occupy M1 and M2 lattice sites whilst Ca$^{2+}$ occupies only M2
sites. Pyroxenoids occur when Ca$^{2+}$ occupies both M1 and M2 sites
(i.e. Ca$^{2+}$ occupancy of M1+M2 $>50$~per~cent), and the chains are
twisted with a repeat of 3 or more. Pyroxenoids are represented here
by the Ca end-member, wollastonite (Ca$_2$Si$_2$O$_6$, Wo$_{100}$).
Mineralogical formulae Wo$_w$En$_x$Fs$_y$ are calculated from the
atomic percent of each of Ca$^{2+}$, En$^{2+}$, and Fe$^{2+}$,
respectively; if no other elements were included $w+x+y\simeq100$.

We obtain spectra from samples listed in Table~\ref{tab:samples}
representing three sides of the pyroxene quadrilateral because these
minerals are abundant in terrestrial environments and in
meteorites. Minerals are named according to the percentage of
wollastonite and enstatite in the unit cell; the ferrosilite content
can be deduced by subtraction or, more accurately, from the chemical
formulae listed in the table.

\begin{enumerate}
\item {\bf The Enstatite (Mg$_2$Si$_2$O$_6$, En$_{100}$) --
  Ferrosilite (Fe$_2$Si$_2$O$_6$, Fs$_{100}$) Series} where Mg$^{2+}$
  is replaced by Fe$^{2+}$. In this orthopyroxene series the crystal
  axes are orthogonal to each other because Fe$^{2+}$ and Mg$^{2+}$
  are of similar size. However, in circumstances of rapid
  crystallisation, one axis can be inclined giving rise to a
  clinoenstatite to clinoferrosilite series with similar compositions,
  but because these are rare in terrestrial samples they are beyond
  the scope of this paper. \citeauthor{Chihara2002} found
  little difference between the infrared spectra of orthoenstatite and
  clinoenstatite at wavelengths shorter than 40~$\mu$m and significant
  differences at longer wavelengths (see Section~\ref{sec:unobs}).\\

\item {\bf The Enstatite--Diopside--Wollastonite (Ca$_2$Si$_2$O$_6$;
  Wo$_{100}$) Clinopyroxene Series} where Mg$^{2+}$ in the M2 site are
  replaced by larger Ca$^{2+}$ cations. Pyroxenes comprise silicates
  with $\leq 50$~per~cent Ca$^{2+}$, thus diopside (CaMgSi$_2$O$_6$;
  Wo$_{50}$En$_{50}$, or Dp$_{100}$) is the end-member pyroxene;
  wollastonite, the end-member pyroxenoid is also included in this
  study because it appears in meteorites. The larger radius of
  Ca$^{2+}$ changes the shape of the unit cell producing crystals with
  an inclined $a$-axis, giving rise to a monoclinic structure
  (clinopyroxenes).\\

\item{\bf The Diopside -- Hedenbergite (CaFeSi$_2$O$_6$;
  Wo$_{50}$En$_{0}$) Clinopyroxene Series} where Fe$^{2+}$
replaces Mg$^{2+}$ in the M1 site; these crystals are also monoclinic.\\

\end{enumerate}

Two pigeonites, whose composition does not fall on the edges of the
pyroxene quadrilateral, are also included to indicate some of the
variety of terrestrial pyroxenes. Pigeonites are monoclinic pyroxenes
with a relatively-low calcium content.

\subsection{Effect of trace elements and impurities in natural samples}
Natural samples tend to contain trace quantities of Cr, Ni, Mn, Ti and
Na which replace $\la 5$ per cent of the main constituents of the
silicates. Although different cations theoretically affect the
spectra, they do not normally do so at this level of substitution
unless the impurity has a substantially different mass, charge or
volume from the cation it replaces. By convention traces of Mn$^{2+}$
and Ni$^{2+}$ are summed with Fe$^{2+}$, because the cations are of
similar size and are indistinguishable from Fe$^{2+}$ in infrared
spectra. Additional samples were measured that have chemical
compositions only slightly different from those reported here, and
their spectra were nearly identical.

Impurities appear as weak bands in regions that are transparent for
the major composition.  We found quartz impurities in our original
En$_1$ and En$_{12}$ spectra and subtracted them from our data
(Appendix~\ref{app:quartz}). Sample En$_{99}$ is a good test case for
less easily distinguishable impurities because the meteoritic sample
is known to contain 5\% of microscopic clinopyroxene inclusions (see
Section~\ref{sec:impurities}), which is the canonical detectable
limit. At higher $\ga 10$ per cent concentrations impurities can also
broaden bands and cause them to blend, the effect of substantial
substitution of Al$^{3+}$ for Si$^{4+}$ will be explored in a later
paper. Here, we report data on samples with the smallest quantities of
minor impurities to focus on effects of the major cation variations (
Mg, Fe, and Ca) and we have found no evidence that impurities
significantly affect the current data set.


\section{Experimental Methods}
\label{sec:expt}
Samples are described in Table~\ref{tab:samples}. The chemical
formulae were computed from published chemical analysis of samples
from the same locality, as cited, or from electron microprobe analysis
performed at Washington University using a JEOL-733 equipped with
Advance Microbeam automation.  The accelerating voltage was 15 kV,
beamcurrent was 30 nA, and beam diameter was 1~$\mu$m. X-ray matrix
corrections were based on a modified \citet{Arm88} CITZAF
routine. Silicates and oxides were used as primary standards.

IR absorption spectra were acquired using an evacuated Bomem DA 3.02
Fourier transform spectrometer with an accuracy of $\thicksim$0.01~\wno.
An SiC source was used for the entire spectral range.  The
number of scans ranged from 500 to 2000.  Far-IR spectra, from
$\thicksim $50 to 650~\wno($\thicksim$200--15~$\mu$m), were obtained
at 1--2~\wno-resolution with an Si-bolometer and, in some cases a
12~$\mu$m coated mylar beamsplitter was used to provide higher throughput
at the lowest frequencies. Mid-IR spectra from $\thicksim $450 to
4000~\wno ($\thicksim$23--2.5~$\mu$m) were obtained with a HgCdTe
detector and a KBr beamsplitter and 1--2-\wno-resolution.

Samples were hand-ground for a maximum of 10 minutes in order to
reduce the effects of crystal orientation without degrading the
long-range crystal structure \citep[see e.g. the study by][for the
  spectroscopic effect of excessive mechanical
  grinding]{Imai2009}. However, pyroxenes tend to form elongated laths
with the long axis parallel to their c-axes (the axes of the chains)
and these tend to lie on their sides when compressed, the crystal
orientation in the DAC will not be perfectly randomized. This effect
will be offset slightly by the non-parallel condensing beam of the
DAC, so that all the crystal axes are sampled.

Optically-thin, $\thicksim $0.2 to 2~$\mu$m-thick, films were made by
compressing powders in a diamond anvil cell (DAC) to form a uniform
thickness by repeatedly applying $\sim 10$ kbar pressure for a few
seconds. Excess pressure is released prior to measuring the spectra so
that the cell acts as a sample-holder (\citeauthor[see][and references
  therein]{HKS2003}) and the crystals revert to their decompressed
state by elastic decompression \citep[see ][for decompression
  after applying hydrostatic pressures of up to $\la$ 425kbar in
  olivine]{Hofmeister1989}.  Various powder thicknesses were used to
confirm the existence of weak peaks, to ascertain that optically-thin
conditions hold for the most intense peaks, and to minimize fringing
in the transparent regions \citep{HB2006}.

Band strengths near 700~\wno were calibrated against a film whose
thickness was determined by using a $1.93~\mu$m-thick microphone foil
as a spacer around an aperture containing a sample which had been
finely ground under alcohol.

Aluminium microphone foils, nominally $2.5\pm0.5~\mu$m- and
4~$\mu$m-thick, were purchased on ebay from Geisnote. Their thickness
was checked by placing a strip of the thicker foil between two
25mm-diameter KBr discs. The absorption spectra showed fringes whose
thickness is given by $B = 1/ (2n \Delta \nu)$, where $\Delta \nu$ is
the average spacing and $n$ is he index of refraction = 1 for the air
gap \citep{Gd1986}. The nominally 4-$\mu$m foil, was found to be
5.3~$\mu$m thick. The thin single foil was below 2~$\mu$m thick because
no fringes were seen. In order to move the interference into the
measurable near--mid-infrared wavelength range, the thinner foil was
doubled and the double-thickness measured to be 3.86~$\mu$m-thick.



\subsection{Data processing}
\begin{figure}
  \includegraphics*[bb=67 180 450 715,  width=\linewidth]{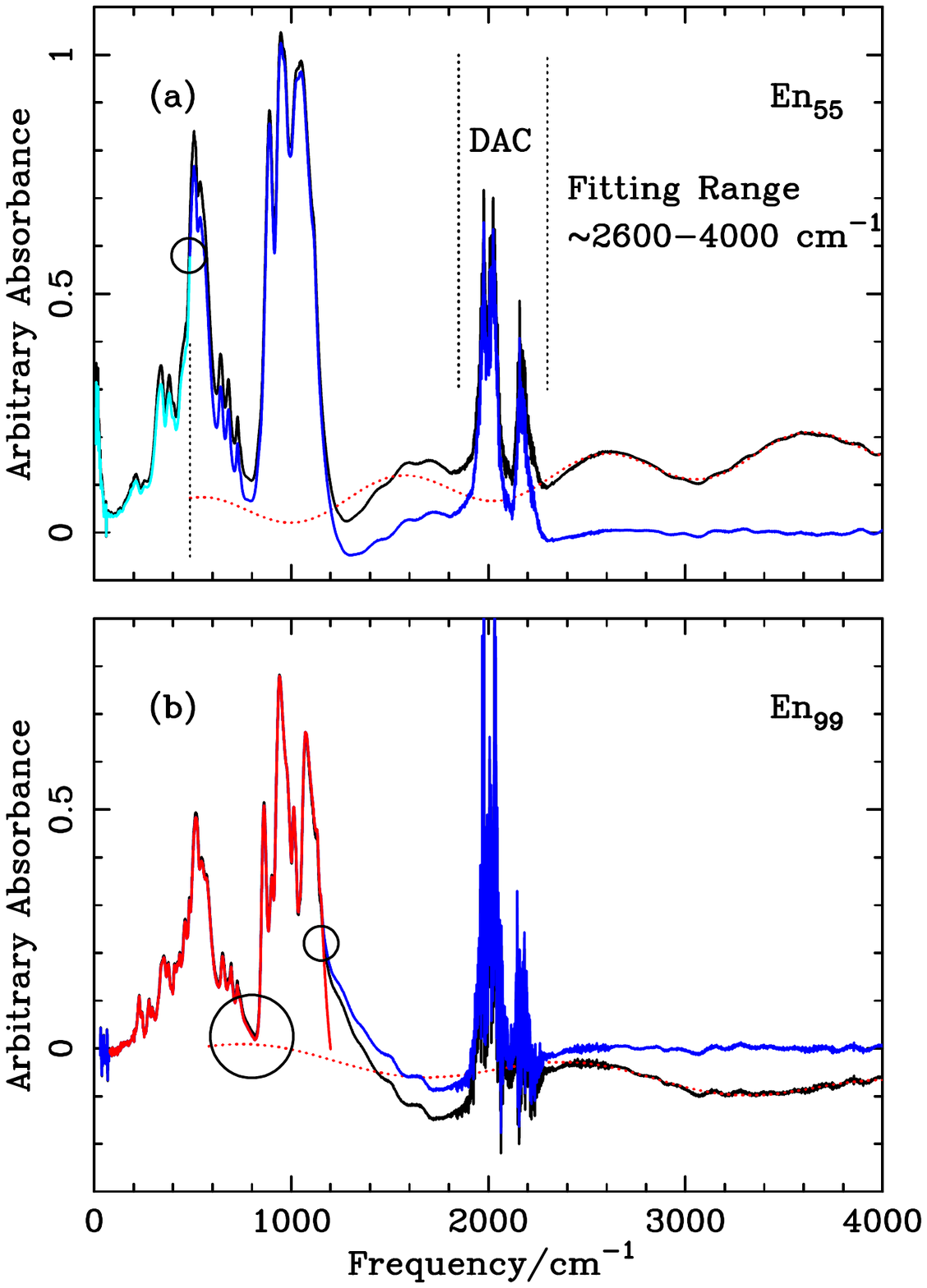}
\caption{(a) The complete spectrum of En$_{55}$ before defringing
  (black) with sinusoidal function fitted to 2600 to 4000 cm$^{-1}$
  -range (red-dotted). This was subtracted as far as the merging point
  near 580~\wno to produce the defringed spectrum. The low
  frequency region (cyan) was obtained by scaling a spectrum of
  another film to match the higher frequency spectrum. DAC denotes the
  region dominated by absorption within the diamand anvil cell. (b)
  Removing the wing near 1100~\wno: black - original spectrum,
  blue- after defringing, red after merging near 1100~\wno. Small
  circles indicate the merging points and the large circle highlights
  the transparent region whose level may be affected by unresolved
  weak peaks.\label{fig:defringe}}
\end{figure}

\begin{figure}
\includegraphics*[bb=66 390 503 790,width=\linewidth]{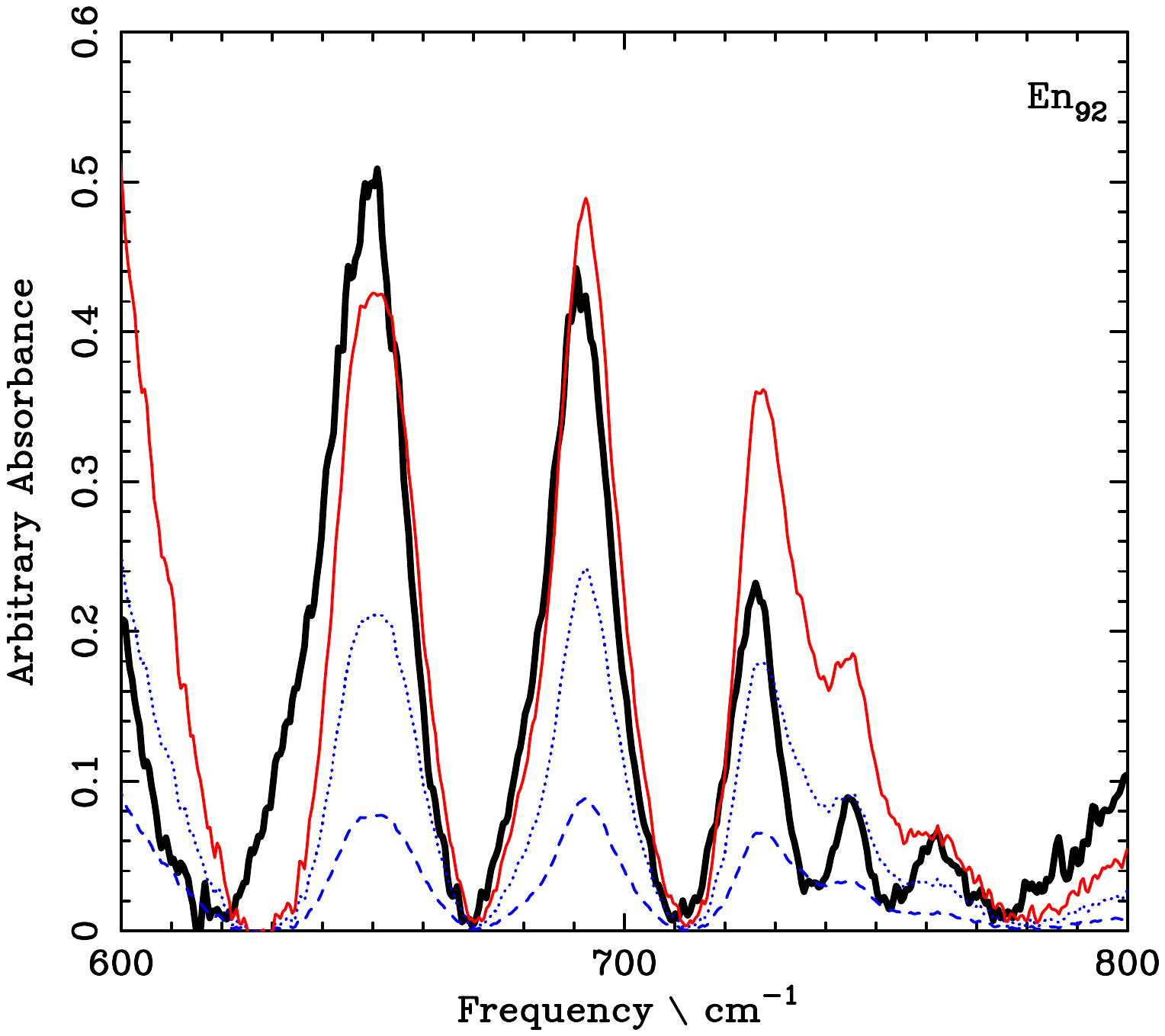}
\caption{Calibrating the absorbance of En$_{92}$. Film of
  known thickness (thick black curve), uncalibrated En$_{92}$ (blue
  dashed), En$_{92}$ matched to $>725$~\wno bands (blue dotted),
  preferred match to $<625$~\wno peaks (solid
  red) \label{fig:bdstr}}
\end{figure}

The 450--4000-\wno spectra contained underlying-fringes which were
removed by fitting and subtracting a sloping sinusoidal function to
the base line at frequencies higher than 2600~\wno
(Figure~\ref{fig:defringe}(a)). Assuming that the absorbance is zero
near zero frequency, the average level of the relatively noisy
80--120-\wno-range was subtracted from the far-IR (50--650~\wno) spectra
and then these were scaled to match the defringed mid-IR data.

The apparent broadening (or wing) of the $\ga1100$-\wno
(9.1-$\mu$m)-region in some samples (Figure~\ref{fig:defringe}(b)) is
not due to absorption. The wing is due to strong reflection at
frequencies above the strongest Si--O stretching (LO) mode
\citep[see][]{Wooten1972}, which reduces the measured transmission,
while enhancing the perceived absorption. If the sample were
negligibly thin, the wing would not exist. This effect is reduced in
films in comparison with the spectra of dispersions, but is not
completely absent because the films are slightly wedged. The wing was
removed by scaling an exponential function modelled on the same,
unaffected, region in En$_{55}$ to the spectra affected and merging it
with the good data at lower frequencies
(Figure~\ref{fig:defringe}(b)). Spectra were subsequently trimmed to
cover the region between 80 and 1200~\wno (125~$\mu$m to
8.3~$\mu$m) and normalised to the area under En$_{55}$ between 80 and
1200~\wno since the area under the spectrum is proportional to the
number of molecular-bonds responsible for the absorption features and
that this should be the same for all the pyroxenes. These are the
normalised laboratory absorbance spectra, ($\bar{a}$, calculated from
$a$ in Equation~\ref{eq:abar} below) used to determine the
compositional dependence of the peak positions.

Uncertainties in the spectral shapes are more significant at the
merging points near 500~\wno (20~$\mu$m), and there is remaining
uncertainty in the intrinsic level of the baseline of the transparent
regions near 80~\wno, 600--800~\wno and 1200~\wno. Additional
measurements of the 600--800~\wno region of En$_{92}$ for band
strength calibration (Figure~\ref{fig:bdstr}), indicate that there may
be unresolved weak peaks near 800~\wno in some of the spectra which
manifest as a change from a U-shape to a V-shape (e.g. in En$_{99}$
and En$_{92}$); they may also be due to the the trace quantities of
other cations listed in Table~\ref{tab:samples}.

\subsection{Band-strength calibration}
\label{sec:bandstr}
Laboratory absorbance
\begin{equation}\label{eq:abar}
a=-\log_{10}\frac{I_t}{I_*}=A \times d,
\end{equation}
where $I_t$ is the intensity of the beam transmitted through the DAC
containing the sample and $I_*$ is the intensity of the beam transmitted
through the empty cell. $a$ is equivalent to the absorption
coefficient, $A$, times the film thickness $d$ in the spectroscopic,
chemical and mineralogical literature. However, astronomers use a
natural log absorption coefficient (or optical depth, $\tau$) units so
our pyroxene data are presented as
\begin{equation}
\label{eq:tau}
  \tau (\mu m^{-1})=\frac{A}{d}\times2.3026,
\end{equation} where the factor of 2.3026 is derived from the change of base formula.

Band strengths for the En--Fs series were estimated by subtracting a
spline fit from the base of the weak peaks between 600--800 ~\wno
region of En$_{92}$ and matching the areas of the two strongest peaks
between 640 and 710~\wno to that of a film of known thickness
(Figure~\ref{fig:bdstr}).

The estimated thickness of the normalised En$_{92}$ spectrum was
0.35~$\mu$m and varying the frequency range of the area calculation
indicated an uncertainty in $d$ of about 0.02~$\mu$m or 6~per~cent. Due to the
prior normalisation to En$_{92}$, the actual thickness of the
mid-infrared-film was $0.29\pm0.02~\mu$m. This procedure was then
repeated for the Wo$_{50}$En$_{50}$ (Dp$_{100}$) sample and a
film-thickness of 0.71~$\mu$m deduced. The natural log absorbance
coefficients, $\tau$, were obtained by scaling the normalised
laboratory-absorbance spectra of the En--Fs orthopyroxenes to the
En$_{92}$ thickness, and the En--Dp--Wo and Dp--Hd clinopyroxenes to
the thickness of Dp$_{100}$.

Values of the calculated mid-infrared sample thicknesses are in
Table~\ref{app:merge}. Knowing the original thickness helps to gauge
possible rounding of the peaks in overly thick
films \citep[e.g.][]{HKS2003}. Each peak in a spectrum has an
associated band strength and thus a range of sample thickness exists
where the band is best observed.

\subsection{Calculation of Mass Extinction Coefficients.}
\label{sec:massext}
We present our data as natural log absorption coefficients because
this represents attenuation and path length. Mass extinction
coefficients, $\kappa$ (cm$^2$g$^{-1}$), can be calculated using the
relation
\begin{equation}
\label{eq:mabs}  
  \kappa=10^4\tau/\rho
\end{equation}
where $\rho$ is the density of the mineral.  Room-temperature
orthopyroxene densities vary in a linear fashion with Mg:Fe ratio
between 3.2~gcm$^{-3}$ (En$_{100}$) and 4.00~gcm$^{-3}$ (En$_0$)
\citep{DHZ1978}. Estimated densities for each of our samples are in
Table~\ref{tab:samples}.

\section{Overview of Spectra}
\label{sec:over}
\begin{figure*}
\includegraphics*[bb=70 50 450 725, width=0.5\linewidth]{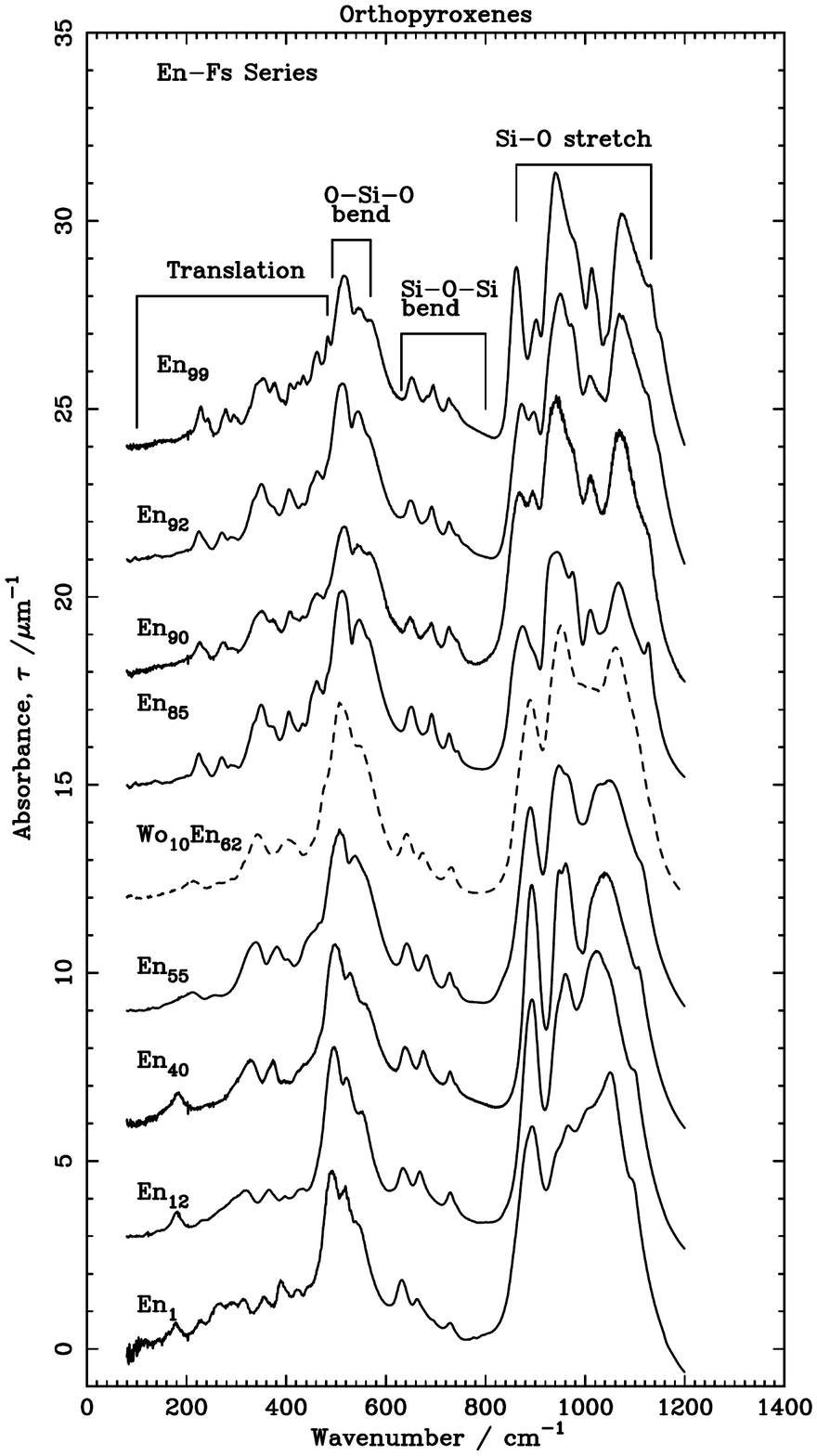}\includegraphics*[bb=70 50 450 725,width=0.5\linewidth]{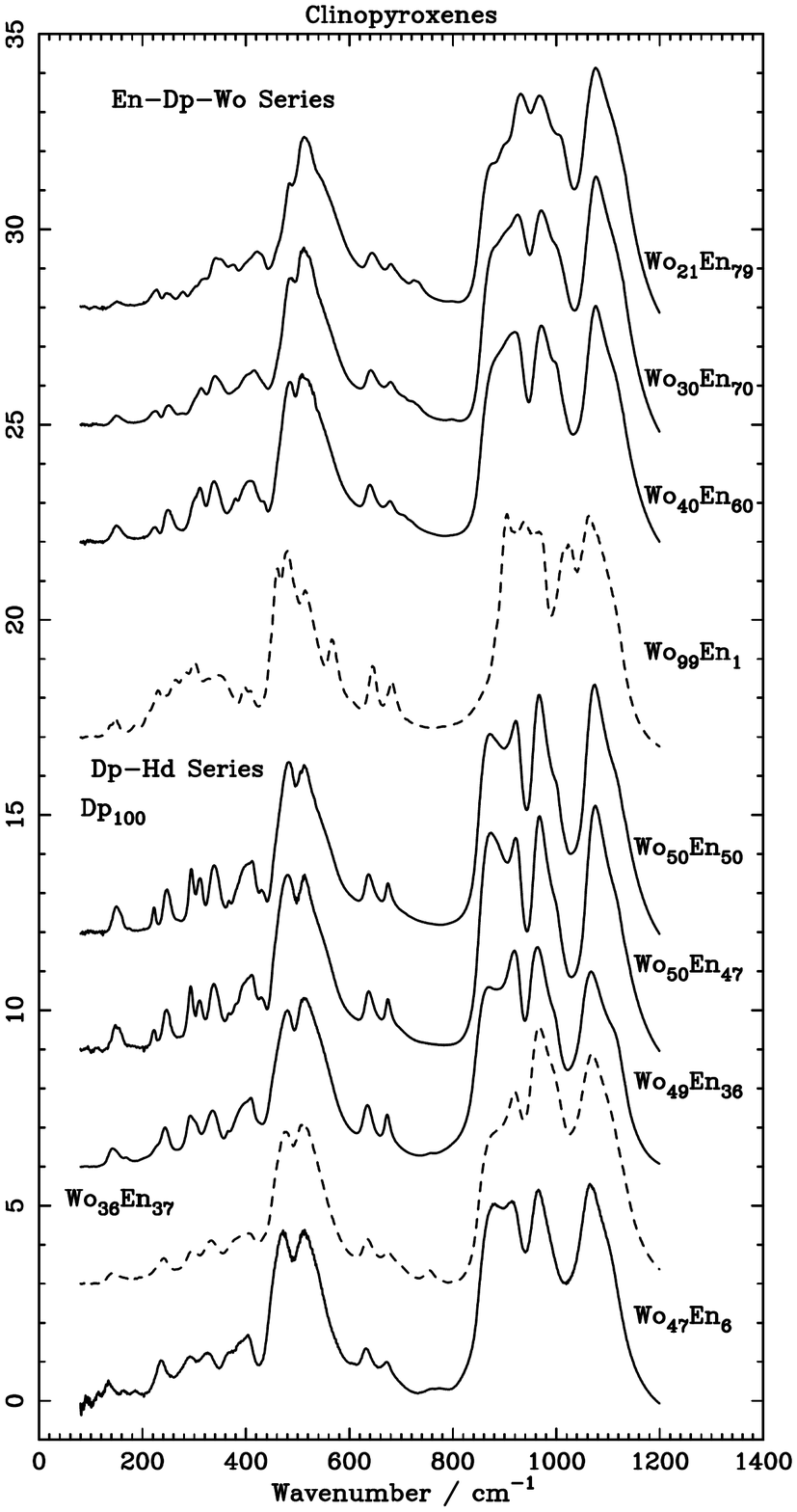}
\caption{Left: Orthopyroxenes in the Enstatite-Ferrosilite
  solid-solution series, each spectrum is offset by 3 from the
  spectrum below it. The spectrum of pigeonite, Wo$_{10}$En$_{62}$
  resembles an En--Fs pyroxene even though it is of intermediate
  composition (dashed curve). Right: Clinopyroxenes in the
  Enstatite-Diopside-Wollastonite and Diopside-Hedenbergite
  solid-solution series. Spectra of the end-member \emph{pyroxenoid}
  wollastonite, Wo$_{99}$ and the Dp--Hd-like spectrum of pigeonite
  Wo$_{36}$En$_{37}$ are dashed because their compositions are not on
  the edges of the pyroxene quadrilateral. Clinopyroxene offsets from
  the bottom are: Dp--Hd: 0, 3, 6, 9, 12; En-Dp-Wo: 17, 22, 25,
  28\label{fig:ortho}.}
\end{figure*}

The complete 80--1200~\wno spectra of the orthopyroxene,
Enstatite--Ferrosilite series and the
Enstatite--Diopside--Wollastonite and Diopside--Hedenbergite Series
are displayed in Figure~\ref{fig:ortho} along with two pigeonites and
end-member wollastonite (dashed). In common with nearly all
low-pressure silicates the spectra have strong Si--O stretching modes
near 1000~\wno (10~$\mu$m) and O--Si--O-bending modes near 500~\wno
due to their SiO$_4$ tetrahedra. Translations due to motions of the
metal cations occur at frequencies below about 450~\wno. Chain
silicates exhibit additional and distinctive weaker bands in the
600--800~\wno- (16.5-- 12.5~$\mu$m) range due to Si--O--Si bending
modes in the linked oxygen atoms of the chains of silicate tetrahedra.

As observed by others \citep[e.g.][]{Jaeger1998, Chihara2002,
  BMG2007}, in the En--Fs series bands generally move to lower
frequencies (longer wavelength, towards the red), as Mg is replaced
with Fe, with larger frequency shifts occurring in the translational
bands. In common with other solid-solution series, endmembers
(En$_{99}$ and En$_{1}$) have more and sharper peaks than intermediate
members. Several bands show dramatic changes when the Fe content
exceeds $\sim 50$~per~cent in behaviour analagous to that previously
observed in the olivine series \citep{HP2007}; this effect will be
discussed in Section~\ref{sec:shifts}.  It is noteworthy that only two
bands are observed to move to \emph{higher} frequencies (to the blue)
with an increase in Fe content, these shifts are small - frequencies
shift between 862--893~\wno (11.6--11.2~$\mu$m) and 460--464~\wno
(21.7--21.51~$\mu$m) and that both shifts occur only at the Mg-rich
($\sim$ En$_{100}$--En$_{40}$) end of the series. We generally
identify similar bands and compositional trends to
\citeauthor{Chihara2002} (see Section~\ref{sec:compare}).

As in the En--Fs series, bands in the Dp--Hd series shift to lower
frequencies as Fe replaces Mg in the monoclinic Dp--Hd series; no
bands shift to higher frequencies with an increase in Fe. Peaks in the
En--Dp--Wo series shift to lower frequencies with replacement of Mg
for Ca because the larger cation forces expansion of the lattice. The
wollastonite spectrum resembles the En--Dp spectra above 400~\wno
because these bands represent nearest-neighbour interactions. But, due
to its different long-range order wollastonite, has a very different
absorption profile in the 200--400-\wno region of the
Translations. Spectra of the pigeonites are similar to pyroxenes on
the nearest edges of the pyroxene-composition quadrilateral (Figure
~\ref{fig:nomenclature}). For example, the effect of 10~per~cent
substitution of Ca for Mg in Wo$_{10}$En$_{62}$ reduces the number of
narrow bands, but does not alter the spectrum much from the rest of
the En--Fs series. In contrast, more substantial replacement of Mg and
Fe by Ca in Wo$_{36}$En$_{37}$ noticeably reduces the height of the
peak near 840~\wno, weakens the translations and adds an extra
Si--O--Si band near 760~\wno in comparison to minerals in the Dp--Hd
Series. Further discussion of pigeonites and wollastonite is beyond
the scope of this paper.

\section{Composition-Dependent Wavelength Shifts}
\label{sec:shifts}
In order to identify minerals in astronomical dust it is necessary to
determine the degeneracy between wavelength-shifts in the spectra due
to temperature and those due to changes in chemical composition. Since
each band arises from part of the crystal structure (e.g. the
Si--O--Si bend or translations of the position of the cations relative
to the tetrahedra) there will be ambiguity between different
silicates, and between silicates and other solids containing similar
bonds. An analogy would be to consider the spectral signature of rings
of carbon atoms in organic molecules: a spectral feature is indicative
of a part of a molecule, not the entire structure.

\begin{figure*}
\begin{minipage}{0.69\linewidth}
\includegraphics*[bb=69 5 524 790, width=0.67\linewidth]{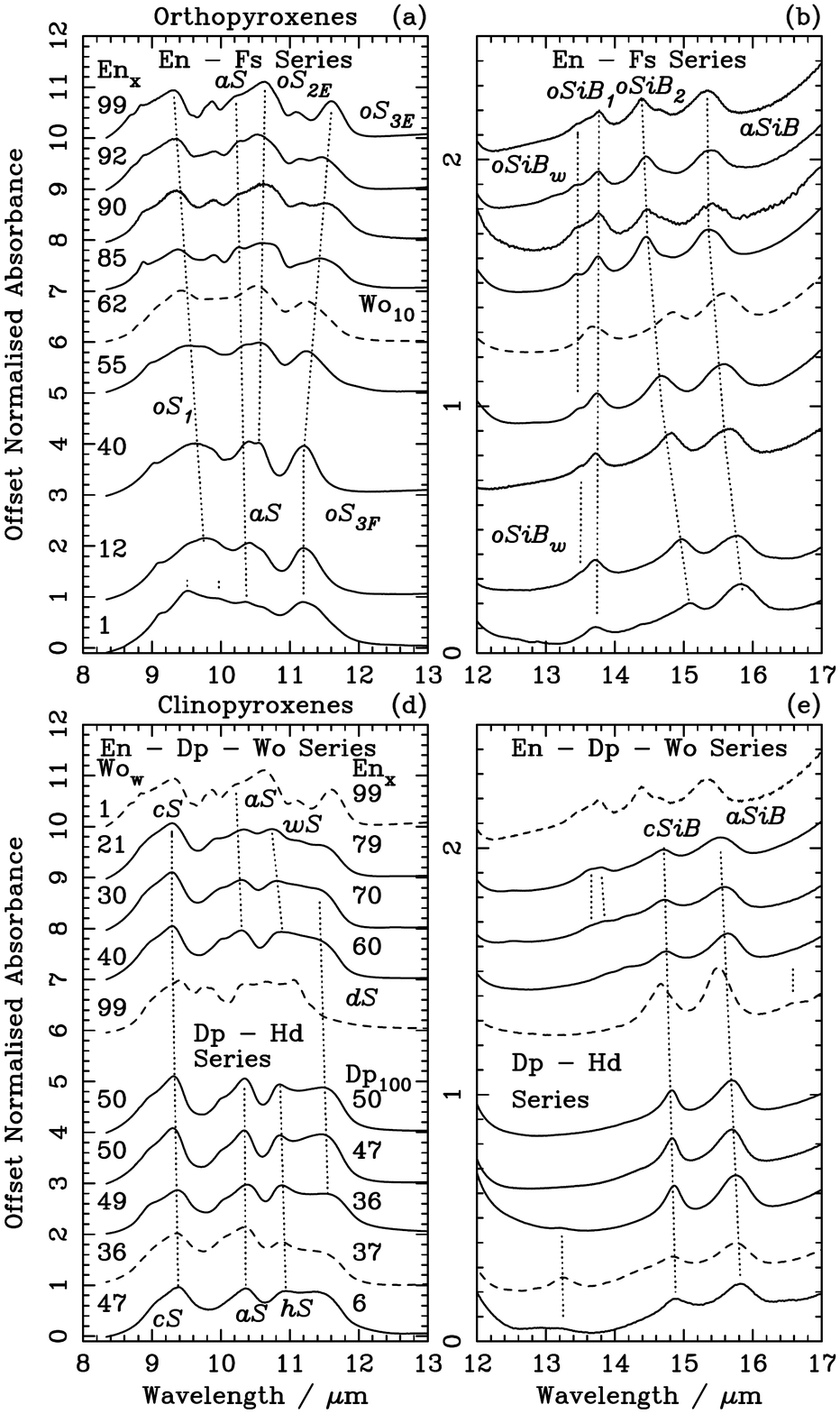}\hfill \includegraphics*[bb=88.5 5 312.5 790, width=0.33\linewidth]{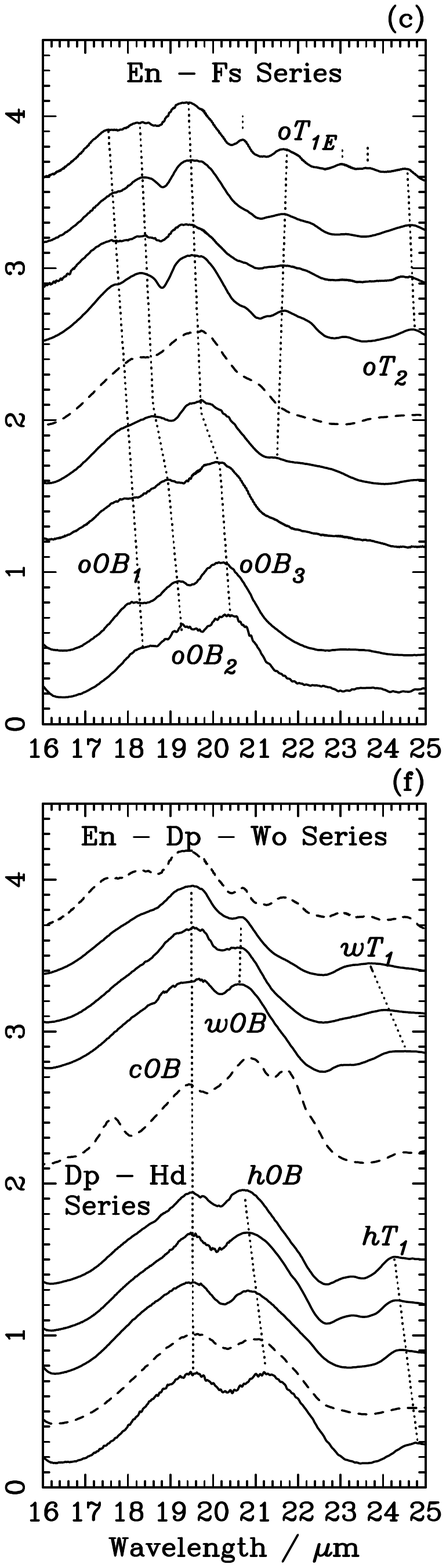}\vfill
\includegraphics*[bb=69 385 524 790, width=0.67\linewidth]{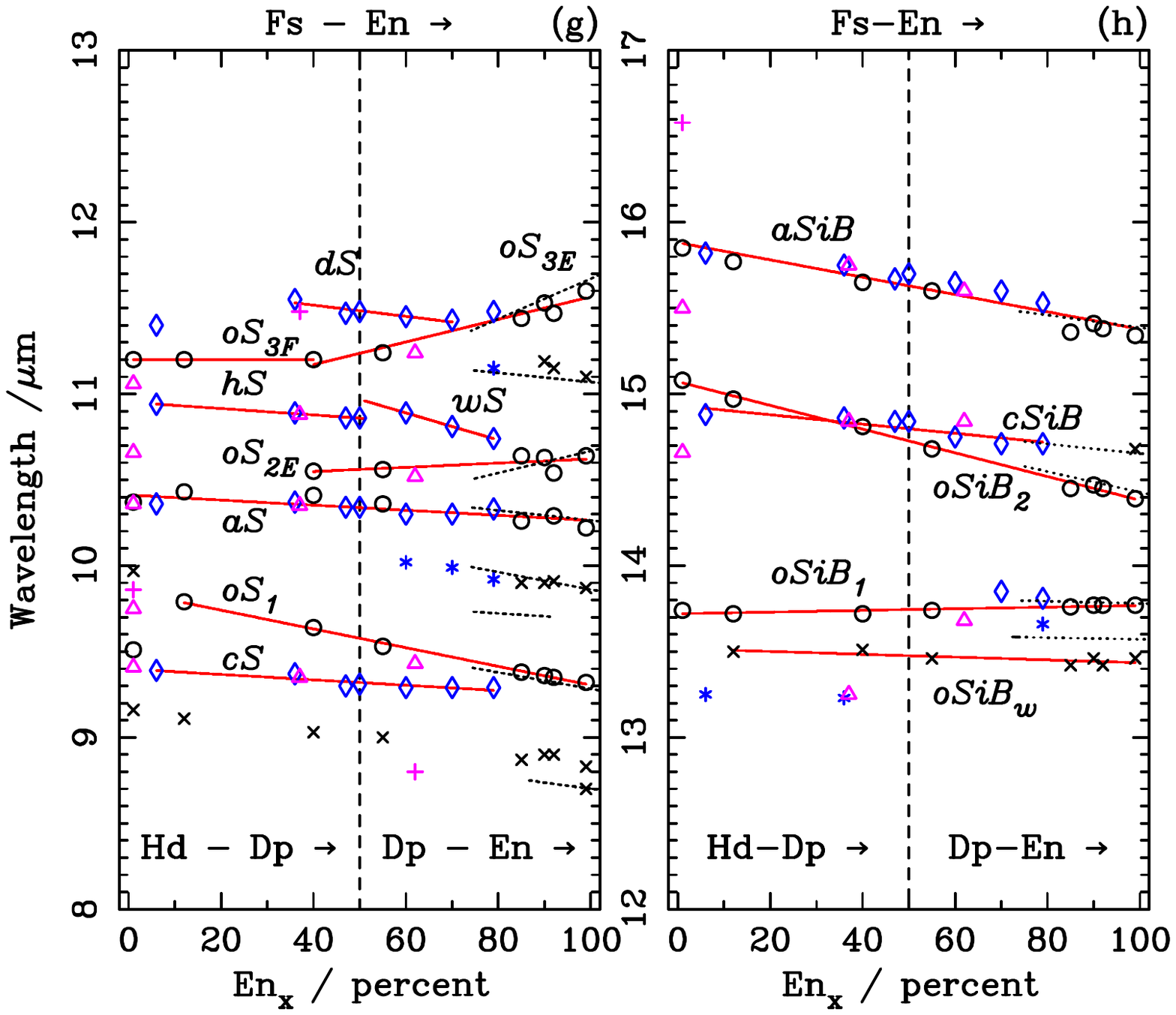}\hfill
\includegraphics*[bb=88.5 385 312.5 790, width=0.33\linewidth]{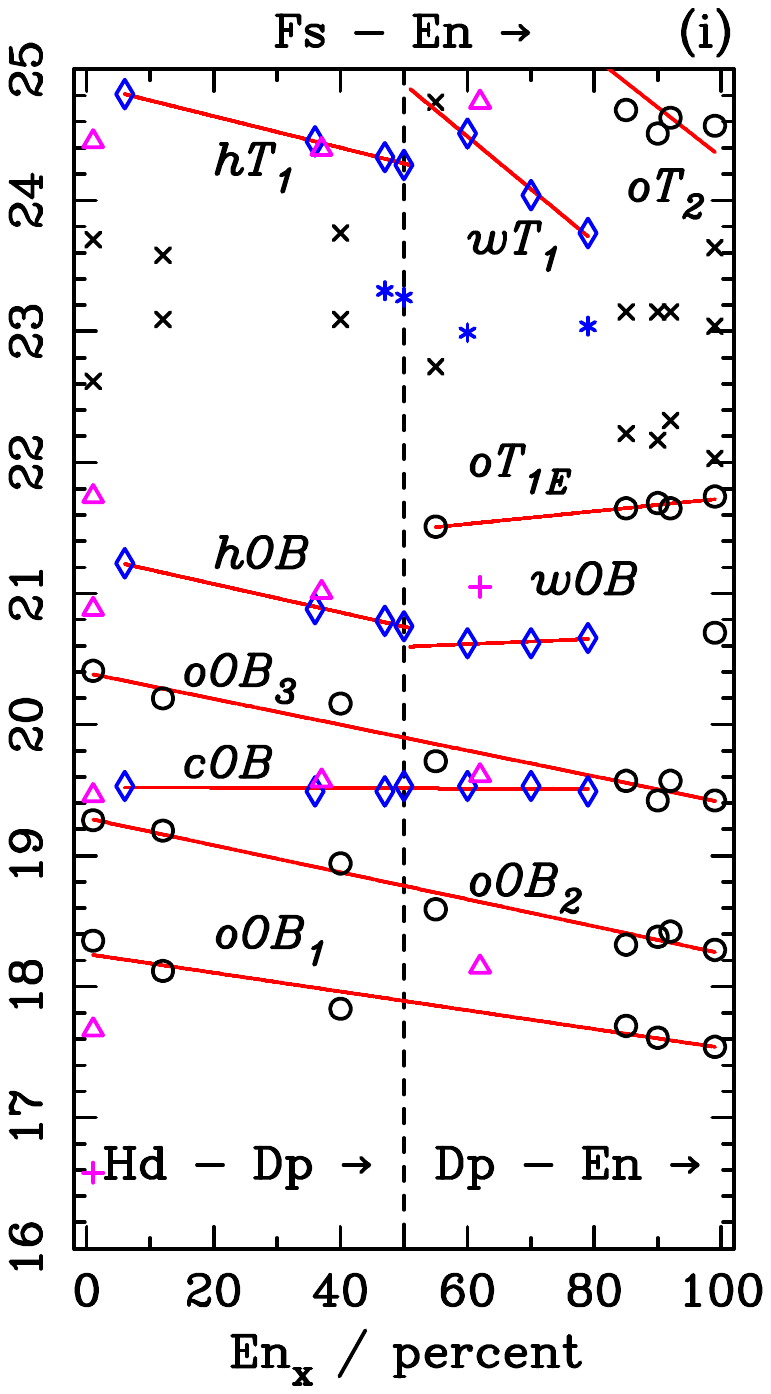}
\caption{Left to Right: Detail in the Si--O stretching, Si--O--Si
  bending modes, O--Si--O bending modes and the highest frequency
  translations of ortho- and clinopyroxenes. Bottom Row: Peak
  positions as a function of Mg content: solid red lines are fits to
  the strongest peak shifts, black dotted lines are fits to the peak
  shifts in meteoritic spectra obtained by ~\citet{BMG2007}. Open
  symbols indicate prominent bands: En--Fs orthopyroxenes -- black
  $\boldsymbol{\medcirc}$, $\times$; En--Dp and Dp--Hd clinopyroxenes
  -- blue $\boldsymbol{\lozenge}$, * and pigeonites and wollastonite
  -- magenta $\boldsymbol{\triangle}$, + \label{fig:sioshifts}}
\end{minipage}
\end{figure*}

Changes in the Si--O stretching, Si--O--Si bending and O--Si--O-bending
modes with composition are presented in Figure~\ref{fig:sioshifts} and
the bands listed in Tables ~\ref{tab:stretchlist} and
\ref{tab:bendlist}; changes in the Translations are examined in
Figures~\ref{fig:sioshifts} and \ref{fig:translations} and the bands
listed in Tables~\ref{tab:bendlist} and \ref{tab:translist}.  The
fitted compositional-dependence of the peak wavelengths of prominent
bands are listed in Table~\ref{tab:peakshift}. Stronger and
distinctive bands are labelled according to their likely carriers:$S$,
Si--O stretch, $SiB$, Si--O--Si bend, $OS$, O--Si--O bend, $T$,
translation. Letters in front of the carrier assignment indicate the
pyroxene series $o$, orthopyroxenes (En--Fs series), $c$,
clinopyroxenes, $h$, Dp--Hd series, $w$, En--Dp--Wo series, $a$, all
pyroxenes studied; the letter $w$ after a band name indicates a weak
band. Some bands of the En--Fs series have different compositional
wavelength shifts at either end of the series, bands at the Mg-rich
end (En$\ga 50$~per~cent) are denoted by the subscript $E$, Fe-rich
bands at the Fe-rich end (En$\la 50$~per~cent) are labeled $F$.
Numerical subscripts increase with increasing wavelength. In the
sections below wavelengths in parathensises are the wavelength of the
pyroxene with the highest Mg content. Due to the complexity of the
pyroxene spectra, band positions were determined by close inspection
of the frequency-based spectra; a fuller treatment of band shapes is
beyond the scope of this paper, but will be presented together with
the structural constraints on a later paper in the mineralogical
literature.

\subsection{Bands common to all studied pyroxenes}

Only two narrow bands are common to all the pyroxenes
(Figure~\ref{fig:sioshifts}, and Tables~\ref{tab:stretchlist},
~\ref{tab:bendlist} and~\ref{tab:peakshift}) an Si--O stretching mode,
$aS$, (10.22~$\mu$m; En$_{99}$ ) and an Si--O--Si-bending mode, $aSiB$
(15.34~$\mu$m; En$_{99}$). $aS$ shifts by only 0.15~$\mu$m as Mg is
replaced; the band is indistinct in the orthopyroxenes, due to the
dominance of a neighbouring blended band ($oS_{2E}$), but becomes more
prominent in the clinopyroxenes when Ca increases from from 20--50
percent, and strongest for diopsides ($\simeq$ Wo$_{50}$En$_{50}$Fs$_{50}$). In
contrast, the common bending-mode, $aSiB$, is prominent across our
range of pyroxene samples and shifts by only 0.51~$\mu$m to
15.85~$\mu$m in En$_1$.

\subsection{Si--O stretches}
\begin{table*}
  \caption{Si--O stretch - peak wavelengths of bands. Pigeonites
    of intermediate composition and bands not used in fitting are
    indicated with bracketed italics. Bands with large compositional
    wavelength shifts are indicated in bold.  Scatter in the
    measurements indicates an uncertainty of $\pm0.015\mu$m \label{tab:stretchlist}}
\begin{minipage}{\linewidth}
\begin{tabular}{lccccccccc}
\hline
Sample&\multicolumn{9}{c}{Band assignment and measured wavelength ($\mu$m)} \\
\hline
      &    &    & $\boldsymbol{oS_1}$&  &  & $aS$ &$oS_{2E}$ & &  $oS_{3E}$\\  
En$_{99}$&8.70&8.83&{\bf  9.32} &  & 9.87 &10.22 & 10.64 & 11.10  & {\bf 11.60}\\
En$_{92}$&    &8.90&{\bf  9.35} &  & 9.91  &10.29 & 10.54     & 11.15 &  {\bf 11.47}\\
En$_{90}$&    &8.90&{\bf  9.36} &  & 9.90  &      & 10.63     & 11.19 &  {\bf 11.53}\\
En$_{85}$&    &8.87&{\bf  9.38} &  & 9.90 &10.26 &  10.64  &    &  {\bf 11.44}\\

({\it Wo}$_{10}${\it En}$_{62}$)&&{\it (8.80)}  &{\it (9.43)}  &&&&{\it (10.52)} &&{\it (11.24)} \\

En$_{55}$&  &9.00&{\bf  9.53} &  &   &10.36 & 10.56 &    &  {\bf 11.24}\\
\cline{10-10}
        &    &    & {\bf      }&    &      &      &       &       &     $oS_{3F}$\\   
En$_{40}$&  &9.03&{\bf  9.64} &  &   &10.41 & 10.55 &    &  11.20\\
En$_{12}$&  &9.11&{\bf  9.79} &  &  &10.43 &     &     &    11.20\\
En$_{1}$ &  &9.16&           &9.51 &9.97   &10.37 &     &    &  11.20\\
\cline{1-5}\cline{8-8}\cline{10-10}
&      &     & cS   &    &      &    &$wS$ &       &    dS \\
Wo$_{21}$En$_{79}$ &      &     &9.29  &    & 9.92 &10.33 &10.74 & 11.15 &   {\it (11.48)}  \\ 
Wo$_{30}$En$_{70}$ &      &     &9.29    &&  9.99  &10.30    &10.81 &     &   11.43  \\   
Wo$_{40}$En$_{60}$ &      &     &9.29  &    & 10.02&10.30 &10.89 &     &   11.45  \\   
\cline{1-3}\cline{6-6}\cline{8-8}

        &    &     &      &     &      &          &$hS$&	    &   \\
Wo$_{50}$En$_{50}$ &      &     &9.31  &    &      &10.34 &10.86 &     &   11.48\\   
Wo$_{50}$En$_{47}$&    &     &9.30  &     &        &10.34 &10.86 &    &   11.47  \\   

({\it Wo}$_{36}${\it En}$_{37}$) &&&{\it (9.35)}&  &&{\it (10.35)} &{\it (10.88)} &&{\it (11.48)}\\

Wo$_{49}$En$_{36}$&    &     &9.37  &     &      &10.37 &10.89 &     &   11.55  \\   
Wo$_{47}$En$_{6}$&    &     &9.39  &     &    &10.36 &10.94 &    &   {\it (11.40)}\\
\hline
\multicolumn{9}{l}{Pyroxenoid (no band assignments)}\\
Wo$_{99}$&&&9.41 & 9.75  &9.86  &10.36 &10.66 &11.06\\
\hline
\end{tabular}
\end{minipage}
\end{table*}

The orthopyroxene bands
(Figure~\ref{fig:sioshifts}(a)+(g), Tables~\ref{tab:stretchlist}
and~\ref{tab:peakshift}) are: $oS_1$, (9.32~$\mu$m; En$_{99}$ ),
$oS_{2E}$ (10.64~$\mu$m; En$_{99}$) and oS$_3$ (11.60~$\mu$m;
En$_{99}$) which is subdivided into two components, oS$_{3E}$ and
oS$_{3F}$. Increases in the proportion of Fe reduce the overall width
of the whole stretching feature due to the shift of $oS_1$ by
+0.47~$\mu$m and the \emph{decrease} in the wavelength of band
$oS_{3E}$ from 11.60~$\mu$m (En$_{99}$) to 11.20~$\mu$m
(En$_{40}$). In contrast, band $oS_{3F}$ remains at 11.2$\mu$m does
not shift in wavelength between En$_{40}$ and En$_1$. Both the
wavelength and width of the band resemble those of peaks seen in many
astronomical sources in which the carrier is frequently identified as
olivine~\citep[e.g. in comets][]{HLR1994} or occasionally as a
polyaromatic-hydrocarbon absorption feature~\citep{Bregman2000}.
It could also contribute to the shoulder identified as olivine
Mg-end-member forsterite \citep[e.g.][]{Do-Duy2020} in young stellar
objects and the ISM. The 10.64--10.55-$\mu$m $oS_{2E}$ band shifts by
only +0.1~$\mu$m between En$_{99}$ and En$_{40}$ and probably merges
with $aS$ at En$\sim 12$~per~cent.

The characteristic Si--O-stretching bands of Ca-bearing clinopyroxenes
(Figure~\ref{fig:sioshifts}(d)+(g), Tables~\ref{tab:stretchlist}
and~\ref{tab:peakshift}) are: $aS$ and $cS$ (9.29$\mu$m ;
Wo$_{21}$En$_{79}$ ). In addition, the En--Dp series has band $wS$
(10.74~$\mu$m; En$_{79}$--En$_{40}$) which shifts by +0.15~$\mu$m and
the Dp--Hd series, $hS$ (10.86~$\mu$m; En$_{50}$--En$_{6}$) shifts by
only +0.08~$\mu$m across the series. Clinopyroxenes with diopside-like
compositions (Wo$_{30}$En$_{70}$ to Wo$_{49}$En$_{36}$) have a
prominent shoulder, $dS$ at 11.43--11.55$\mu$m. In common with the
orthopyroxenes, the width of the overall Dp--Hd series Si--O
stretching region becomes narrower with increasing Fe content.

Wavelength shifts of $aS$, $cS$ and $hS$ have a similar dependence on
the percentage of Mg $\sim 0.002\times x$, where $x$ is the percentage
of Mg (Figure~\ref{fig:sioshifts}(g), Table~\ref{tab:peakshift}).

\subsection{Si--O--Si bend}

\begin{table*}
\caption{O--Si--O, Si--O--Si Bends and the highest frequency orthopyroxene translation, uncertainty in wavelength $\sim \pm 0.02\mu$m near 13~$\mu$m and $\sim \pm 0.04\mu$m at 20~$\mu$m \label{tab:bendlist}}
\begin{minipage}{\linewidth}
\begin{tabular}{lccccccccccccc}
\hline
Sample&\multicolumn{8}{c}{Band assignment and peak wavelength ($\mu$m)} \\
\hline                                                    
&           & $oSiB_w$&$oSiB_1$&  $\boldsymbol{oSiB_2}$&            & $\boldsymbol{aSiB}$    & $\boldsymbol{oOB_1}$& $\boldsymbol{oOB_2}$ &$\boldsymbol{oOB_3}$&&$oT_{1E}$ &  \\
En$_{99}$       &  & 13.46&  13.77& {\bf 14.39}&  {\it 14.68} & {\bf 15.34}  & {\bf  17.54}& {\bf 18.28}&{\bf  19.42} &  {\it 20.70}&21.74 & 23.04\\
En$_{92}$       &   & 13.42&  13.77& {\bf 14.45}& &{\bf 15.38}  &              & {\bf 18.42}&{\bf  19.57} &       &21.65 & 23.15 \\
En$_{90}$       &     & 13.46& 13.77& {\bf 14.47}&      & {\bf 15.41}  & {\bf  17.61}& {\bf 18.38}&{\bf  19.42} &       &21.69 & 23.15 \\
En$_{85}$       &     & 13.42& 13.76& {\bf 14.45}&      & {\bf 15.36}  & {\bf   17.70}& {\bf 18.32}&{\bf  19.57}&       &21.65 & 23.15 \\
\cline{13-13}

({\it Wo}$_{10}${\it En}$_{62} $)&&&{\it (13.68)}&({\it 14.84}) &     &{\it (15.60)}  &              &{\it (18.15)} &{\it (19.61)}&&{\it (21.05)}\\

En$_{55}$       &     & 13.46& 13.74& {\bf 14.68}&     & {\bf 15.60}  &      & {\bf 18.59}&{\bf  19.72} &   &21.51 \\
\\                                                             
En$_{40}$       &     & 13.51&  13.72& {\bf 14.81}&     & {\bf 15.65}  & {\bf  17.83}& {\bf 18.94}&{\bf  20.16} & &     & 23.09     \\
En$_{12}$&  & 13.50& 13.72& {\bf 14.97}&     & {\bf 15.77}  & {\bf  18.12}& {\bf 19.19}&{\bf  20.20} & &    & 23.09     \\
En$_{1}$   &   &   &  13.74& {\bf 15.08}&      & {\bf 15.85}  & {\bf  18.35}& {\bf 19.27}&{\bf  20.41} & & & 22.62     \\
\cline{1-3}\cline{5-5}\cline{8-10}\cline{12-12}
                 &         &       &     &   &$cSiB$& &               &&$cOB$   &  $wOB$&    &  \\
Wo$_{21}$En$_{79}$      &      & 13.66 &{\it 13.81}& &14.71 &  {\bf 15.53}      &   && 19.49&  20.66&   & 23.04\\
Wo$_{30}$En$_{70}$      &      &       &{\it 13.85}& &14.71 &  {\bf 15.60}      &   &&  19.53&  20.62&  &      \\
Wo$_{40}$En$_{60}$      &      &       &       &    &14.75 & {\bf 15.65}  &   && 19.53&  20.62&   & 22.99\\
\cline{1-4}\cline{11-11}
                      &       &       &     &      &      &        &     &&      &$hOB$   & &  23.31\\
Wo$_{50}$En$_{50}$      &      &       &       &  &14.84 & {\bf 15.70}  &   && 19.53&  20.75&   & 23.26\\
Wo$_{50}$En$_{47}$      &    &       &     &      &14.84 & {\bf 15.67}  &    && 19.49&  20.79 & \\
Wo$_{49}$En$_{36}$      &13.23 &       &     &      &14.86 & {\bf 15.75}  &    && 19.49&  20.88 & \\
({\it Wo}$_{36}${\it En}$_{37}$)&{\it (13.25)}&&&&{\it (14.84)} &{\it (15.75)} &&&{\it (19.57)}&{\it (21.01)}&\\
Wo$_{47}$En$_{6}$       &13.25 &       &     &      &14.88 &{\bf 15.82}  &    && 19.53&  21.23&  \\
\hline
\multicolumn{9}{l}{Pyroxenoid bands (no band assignments)}\\
Wo$_{99}$&&&&&14.66 &15.50 &16.58 &17.67 &19.46 &20.88&21.74  \\

\hline
\end{tabular}
\end{minipage}
\end{table*}

These modes are produced in the linkages between SiO$_4$ tetrahedra;
we observed three prominent bands in orthopyroxenes
(Figure~\ref{fig:sioshifts} (b)+(h)), but only two bands in
clinopyroxenes (Figure~\ref{fig:sioshifts}(e)+(h));
Tables~\ref{tab:bendlist} and~\ref{tab:peakshift}.
The three
orthopyroxene-bands are $aSiB$ (15.34~$\mu$m; En$_{99}$),
$oSiB_1$ (13.77~$\mu$m; En$_{99}$) and $oSiB_2$ (14.39~$\mu$m; En$_{99}$);
there is also a weak band $oSiB_w$ (13.46~$\mu$m; En$_{99}$)
in En$_{99}$ to En$_{12}$. $oSiB_w$ and $oSiB_1$ barely
shift with composition ($\leq 0.04~\mu$m) but $oSiB_2$ shifts by
0.69~$\mu$m.  The clinopyroxenes and pyroxenoid have two Si--O--Si
bands: $aSiB$ and $cSiB$ (14.71~$\mu$m in Wo$_{21}$En$_{79}$); $cSiB$
shifts by 0.17~$\mu$m across the range to Wo$_{47}$En$_{6}$.

The Si--O--Si bending-modes might be the most promising region for the
identification of chain silicates in the 5--28-$\mu$m spectral-range
of the \emph{Mid-Infrared Instrument (MIRI)} on the
\emph{James Webb Space Telescope (JWST)}
because the spectra of other silicate groups
previously studied by us are quite different between 12 and 16~$\mu$m:
olivines do not display these bands due to the absence of tetrahedral
linkages \citep[e.g.][]{HP2007} and of the hydrous silicates, talc has
only one narrow band at 14.92~$\mu$m which might overlap with some of
the diopsides, and amphiboles have only a single band at 13.15~$\mu$m
\citep{HB2006}; silicas (SiO$_2$ minerals) have narrow bands at
12.5--12.8~$\mu$m, 14.4--14.6~$\mu$m \citep{Koike2013} which are
different from the pyroxene bands.

\subsection{O--Si--O bends}
For this motion, orthopyroxenes have three narrow bands, whereas
clinopyroxenes have two broader bands.

The orthopyroxene bands (Figure~\ref{fig:sioshifts}(c)+(i),
Tables~\ref{tab:bendlist} and~\ref{tab:peakshift}) are:
$oOB_1$~(17.54~$\mu$m; En$_{99}$) which shifts by 0.81~$\mu$m between
En$_{99}$ and En$_1$ and $oOB_2$ (18.28$~\mu$m; En$_{99}$) and $oOB_3$
(19.42~$\mu$m; En$_{99}$) which both shift by 0.99~$\mu$m. The
behaviour of the clinopyroxene bands
(Figure~\ref{fig:sioshifts}(f)+(i), Tables~\ref{tab:bendlist}
and~\ref{tab:peakshift}) is different: the common $cOB$ band
(19.49~$\mu$m; Wo$_{21}$En$_{79}$) does not shift significantly with
composition, the Dp--Hd-band, $hOB$ (20.75~$\mu$m; Dp$_{100}$) shifts
by 0.48~$\mu$m across the range En$_{50}$--En$_{6}$ whilst the
En--Dp-band $wOB$ \emph{decreases} in wavelength by 0.44~$\mu$m as Ca
replaces Mg (En$_{79}$ to En$_{60}$). The non-shifting $cOB$-band
overlaps $oOB_3$ above En$_{90}$: so a similarly-shaped feature at
20.0--20.4~$\mu$m could indicate an orthpyroxene with En$\la
60$~per~cent.

\subsection{Translations}
\begin{figure*}
\begin{minipage}{0.75\linewidth}
\includegraphics*[bb=69 5 524 790, width=0.67\linewidth]{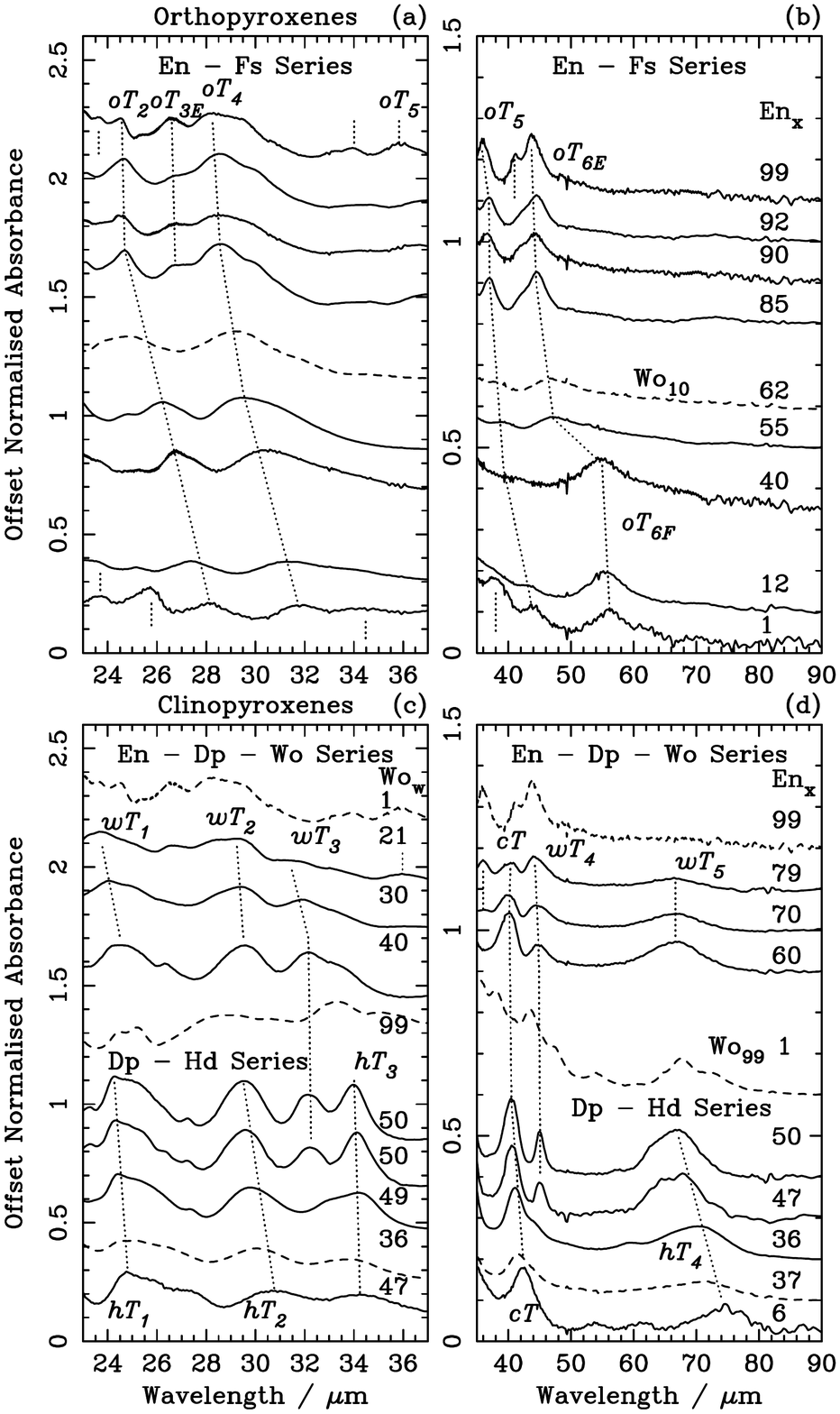}\vfill\includegraphics*[bb=69 385 524 790, width=0.67\linewidth]{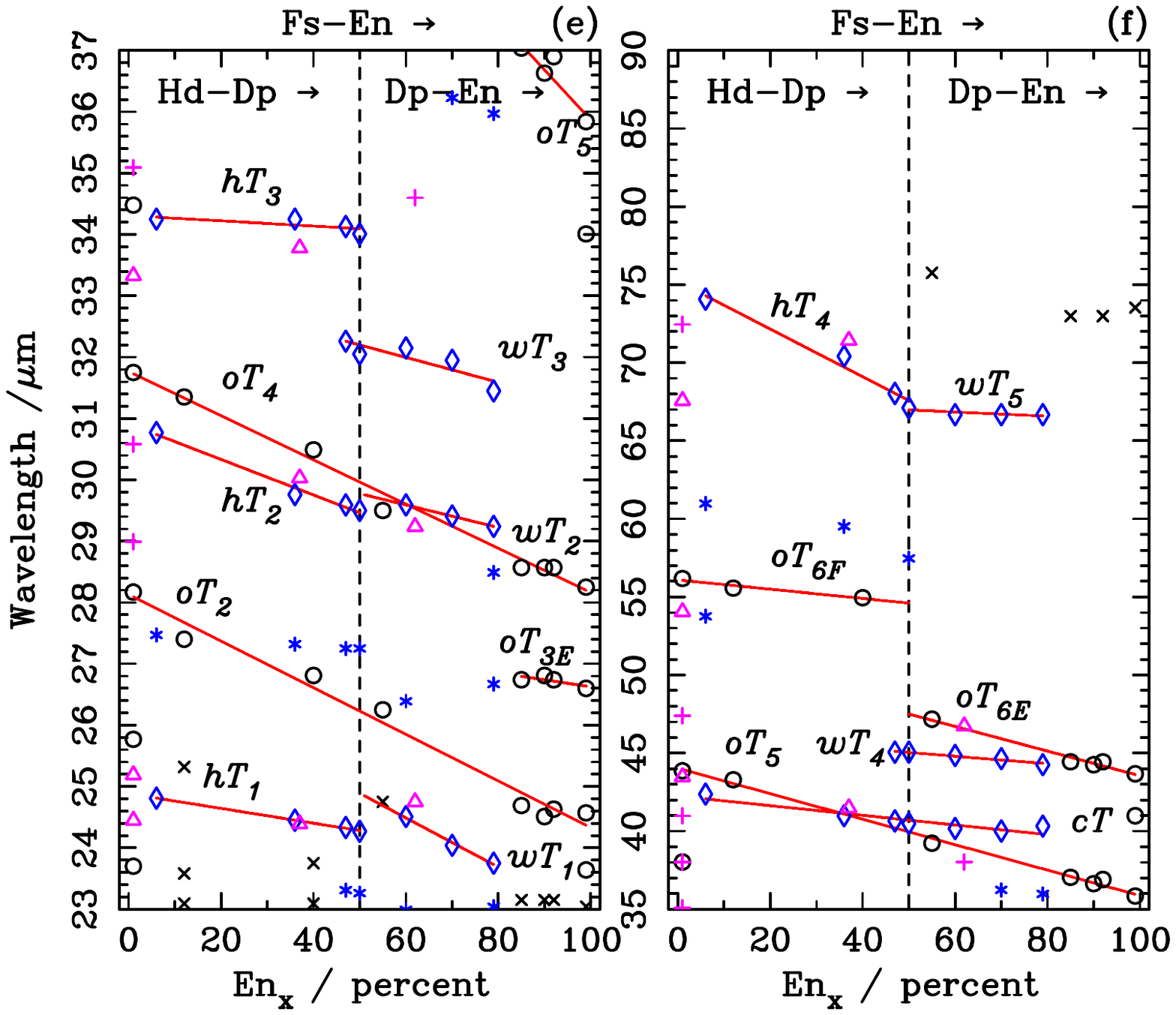}
\caption{Detail in the Translational
  bands\label{fig:translations}. Bottom Row: Peak positions as a
  function of Mg content. See Figure~\ref{fig:sioshifts} for key to plotting symbols.}
\end{minipage}
\end{figure*}
\begin{table*}
\caption{Translations.  Uncertainty at 20~$\mu$m $\pm 0.04$, for the broad clinopyroxene 60$\mu$m peak is $\pm 0.5~\mu$m. The orthopyroxene peaks above 70$\mu$m are ill-defined in our data (see Section~\ref{sec:unobs})\label{tab:translist}}
\begin{minipage}{\linewidth}
\begin{tabular}{lcccccccccccccc}

\hline
Sample&\multicolumn{14}{c}{Band assignment and peak wavelength ($\mu$m)} \\
\hline
                  &   &          &$\boldsymbol{oT_2}$ &$oT_{3E}$   &$\boldsymbol{oT_4}$    &        &       &&$\boldsymbol{oT_5}$    &       &$\boldsymbol{oT_{6E}}$ &&?\\
En$_{99}$&  23.64 &        &{\bf 24.57} &26.60  &{\bf 28.25}   &29.50&34.01\footnote{merged with weak band at 33.11, very broad weak peaks are seen in other members of the group} &       &{\bf 35.84}  &40.98  &{\bf 43.67} &&73.5\\
En$_{92}$   &    &        &{\bf 24.63} &26.74  &{\bf 28.57}   &30.12&      &       &{\bf 36.90}  &       &{\bf 44.44} &&73.0\\
En$_{90}$ &     &        &{\bf 24.51} &26.81  &{\bf 28.57}  &&      &        &{\bf 36.63}  &       &{\bf 44.25} &\\
En$_{85}$ &     &        &{\bf 24.69} &26.74  &{\bf 28.57}   &30.03&      &       &{\bf 37.04}  &       &{\bf 44.44} &&73.0\\
({\it Wo}$_{10}${\it En}$_{62} $)&&&{\it (24.75)}& &{\it (29.24)}& &&&{\it (38.02)} &&{\it (46.73)} \\
En$_{55}$   &     & 24.75       &{\bf 26.25} &    &{\bf 29.50}  &&      &        &{\bf 39.22}  &       &{\bf 47.17}      \\
                     &        &          &            &       &          &         &          &       & &&      &$\boldsymbol{oT_{6F}}$\\
En$_{40}$       &  23.75    &       &{\bf 26.81} &     &{\bf 30.49}&     &      &        &{\bf    }    &    &   &{\bf 54.95} &75.8\\
En$_{12}$       &  23.58    & 25.32       &{\bf 27.40} &     &{\bf 31.35}&   &      &        &{\bf 43.29}  &     &  &{\bf 55.56} \\
En$_{1}$        &  23.70   & 25.77    &{\bf 28.17} &     &{\bf 31.75}&     &34.48    &38.02   &{\bf 43.86}  &     &  &{\bf 56.18} \\
\hline
                         & $wT_1$&     &       &&&$wT_2$ &$wT_3$     &      &&$\boldsymbol{cT}$     &$wT_4$    &    &$wT_5$  \\   
Wo$_{21}$En$_{79}$  & 23.75&&&26.67  &28.49  &29.24  &31.45      & &35.97&{\bf 40.32}   &44.25      && 66.7     \\
Wo$_{30}$En$_{70}$  &  24.04&  &&&       &29.41  &31.95      &      &  &{\bf 40.00}   &44.64      && 66.7     \\
Wo$_{40}$En$_{60}$  & 24.51&&26.39&  &       &29.59  &32.15      &      &  &{\bf 40.16}   &44.84      && 66.7     \\
\hline
& $hT_1$ &       &&       &$\boldsymbol{hT_2}$ &            &$hT_3$ &    &        &                &&& $\boldsymbol{hT_4}$  \\   

Wo$_{50}$En$_{50}$ & 24.27&  &&27.25  &{\bf 29.50}  &32.05         &34.01    &&   &{\bf 40.49}   &45.05     && {\bf 67.1}    \\
Wo$_{50}$En$_{47}$ &24.33 &     &&27.25  &{\bf 29.59}  &32.26         &34.13    &&   &{\bf 40.65}   &45.05     && {\bf 68.0}  \\
Wo$_{49}$En$_{36}$ & 24.45&     &&27.32  &{\bf 29.76}  &              &34.25    &&   &{\bf 40.98}   &          && {\bf 70.4}  \\
({\it Wo}$_{36}${\it En}$_{37}$)         &{\it (24.39)}&{\it (25.19)} &{\it (26.53)} &&{\it (30.03)} &&{\it (33.78)}&& &{\it (41.49)} & &&{\it (71.43)} \\

Wo$_{47}$En$_{6}$  & 24.81&    &&27.47  &{\bf 30.77}  &              &34.25    &&   &{\bf 42.37}   &          && {\bf 74.1}  \\
\hline
\multicolumn{10}{l}{Pyroxenoid bands (no band assignments)}\\
Wo$_{99}$&24.45&25.19&&&28.99 & 33.33& 35.09& 38.02& 43.48&& 47.39& 54.05& 67.6&  72.5\\
\hline
\end{tabular}
\end{minipage}
\end{table*}
Translational bands exhibit the largest wavelength shifts with
composition because cations are less tightly bound than are the chains
of SiO$_4$ tetrahedra. Two orthopyroxene translations (
Figures~\ref{fig:sioshifts}(c),(f),(i) and
~\ref{fig:translations}(a),(b),(e),(f)) have small shifts in the
En$_{99}$--En$_{55}$-range: $oT_{1E}$ (21.74~$\mu$m; En$_{99}$)
\emph{decreases} in wavelength by 0.23~$\mu$m, while and $oT_{3E}$
(26.60~$\mu$m; En$_{99}$) moves by +0.14~$\mu$m.  In contrast, bands
$oT_2$ (24.57~$\mu$m; En$_{99}$) and $oT_4$ (28.25~$\mu$m; En$_{99}$)
each shift by $\sim3.5~\mu$m across the En$_{99}$--En$_1$-range.
$oT_5$ (35.84~$\mu$m; En$_{99}$) has the largest shift of any pyroxene
band: 8.02~$\mu$m En$_{99}$--En$_{1}$. \citeauthor{Chihara2002} found the
largest shift was for a band at 43~$\mu$m in En$_{100}$ which moved to
56~$\mu$m in En$_0$. With our wider range of orthopyroxenes, we find
that this band has two disjointed components: $oT_{6E}$ which shifts
from 43.7~$\mu$m (En$_{99}$) to 47.2~$\mu$m (En$_{55}$) and $oT_{6F}$
which occurs at 55.0~$\mu$m (En$_{40}$) and 56.2~$\mu$m (En$_1$)
giving a combined shift of 12.5~$\mu$m.

The clinopyroxene-series' (Figure~\ref{fig:translations}(c),(d)) have
only one common translation, $cT$, (40.32~$\mu$m ; Wo$_{21}$En$_{79}$)
which shifts by 2.05~$\mu$m across the En$_{79}$--En$_6$ range. The
shifts of En--Dp pyroxenes are relatively small: $wT_1$(23.75~$\mu$m;
Wo$_{21}$En$_{79}$) and $wT_2$ (29.24~$\mu$m; Wo$_{21}$En$_{79}$)
shift in the En$_{79}$--En$_{60}$ range, by 0.76 and 0.35~$\mu$m,
respectively. $wT_2$ shifts in parallel with $wT_3$ (31.45~$\mu$m;
Wo$_{21}$En$_{79}$) and $wT_4$ (44.25~$\mu$m; Wo$_{21}$En$_{79}$ )
which vary by $\sim 0.002\times x$ down to $x=47$. $wT_5$
(66.7~$\mu$m) is the broadest band measured ($\sim 10\mu$m), but
shifts by only 0.44~$\mu$m despite its long peak-wavelength.

In common with the orthopyroxenes, Dp--Hd clinopyroxenes show much
more variable shifts: $hT_1$ (24.27~$\mu$m; Dp$_{100}$) and $hT_3$
(34.01~$\mu$m; Dp$_{100}$) shift by 0.54 and 0.24~$\mu$m, respectively
$hT_2$ (29.5~$\mu$m; Dp$_{100}$) has a similar slope to $cT$, and the
total shift is 1.27~$\mu$m. However, the most striking difference in
the clinopyroxene shifts is that in contrast to the $wT_5$ band of
En--Dp clinopyroxenes, the broadest and lowest frequency band of the
Dp--Hd series $hT_4$ (67.1~$\mu$m; Dp$_{100}$) has a large wavelength
shift of 7.0~$\mu$m between En$_{50}$ and En$_6$.

\begin{table*}
\caption{Dependance of the peak positions of stronger and
    distinctive bands on the proportion of Mg in the pyroxenes.Bands
    with the largest wavelength shifts are highlighted in bold, those
    with negligible wavelength shifts, or small wavelength shifts with
    low Pearson correlation coefficients $|R^2|<0.7$ are
    italicised.\label{tab:peakshift}}.
\begin{minipage}{\linewidth}
\begin{tabular}{llllrll}
\hline
 Series&Label\footnote{Bands are labelled according to
  their likely carriers:$S$, Si--O stretch, $SiB$, Si--O--Si bend, $OS$,
  O--Si--O bend, $T$, translation. Letters in front of the carrier
  assignment indicate the pyroxene series $o$, orthopyroxenes (En--Fs
  series), $c$, clinopyroxenes, $h$, Dp--Hd series, $w$, En--Dp--Wo
  series, $a$, all pyroxenes studied; the letter $w$ after a band name
  indicates a weak band. Some bands of the En--Fs series have
  different compositional wavelength shifts at the En and Fs ends of
  the series, bands at the Mg-rich-end (En $\ga 50$~per~cent) are
  denoted by the additional subscript $E$, bands at the Fe-rich end
  (En$\la 40$~per~cent) are denoted with $F$. Numerical subscripts
  increase with increasing wavelength.}&$\lambda$&Range&$\Delta\lambda$\footnote{$\Delta\lambda=\lambda(En_{max})-\lambda($En$_{min})$, where En$_{max}$ and En$_{min}$ are the maximum and minimum per cent fractions, respectively.}&Fit&$|R^2|$\\
&&$En_{max}$&~per~cent&$\mu$m&En$_0\;\;\; +\; b$En$_x$&\\
 \hline
\multicolumn{4}{l}{\bf Si--O stretch}\\
All  &$aS$ &10.22&99--~~1 &0.15&$10.412-0.0015\,$ &0.8    \\
En--Fs&$\boldsymbol{oS_1}$ &{\bf ~~9.32}&{\bf 99--12}&{\bf 0.47}&$\boldsymbol{ ~~9.850 -0.0055\,}$ &{\bf 1}      \\
     &$\mathit{oS_{2E}}$ &{\it 10.64}&{\it 99--40}&{\it 0.09}&$\mathit{10.500+0.0012\,}$ &{\it 0.6}    \\
     &$\boldsymbol{oS_{3E}}$  &{\bf 11.60}&{\bf 99--40}&{\bf --0.40}&$\boldsymbol{10.908+0.0066\,}$&{\bf 1}\\     
     &$\mathit{oS_{3F}}$  & {\it 11.20}&{\it 40--~~1}&{\it 0.00}&{\it 11.20}&{\it --}\\      
Clino&$cS$&~~9.29  &79--~~6  &0.10&$~~9.398-0.0016\,$     &0.9    \\
     &$dS$ &11.43 &70--36 &0.12&$11.648-0.0033\,$    &0.9    \\   
En--Wo&$wS$   &10.74 &79--40 &0.15&$11.364-0.0079\,$     &1      \\
Dp--Hd&$hS$   &10.86 &50--~~6  &0.08&$10.952-0.0019\,$    &1      \\
\hline                                           
\multicolumn{4}{l}{\bf Si--O--Si bend}\\         
All  &$\boldsymbol{aSiB}$  &{\bf 15.34}&{\bf 99--~~1}  &{\bf 0.51}&$\boldsymbol{15.882-0.0050\,}$   &{\bf 0.9}    \\
En--Fs&$\mathit{oSiB_w}$&{\it 13.46}&{\it 99--12} &{\it 0.04}&$\mathit{13.517-0.0008\,}$&{\it 0.8}   \\
     &$\mathit{oSiB_1}$&{\it 13.77}&{\it 99--~~1}  &{\it -0.03}&$\mathit{13.720+0.0005\,}$   &{\it 0.9}   \\
     &$\boldsymbol{oSiB_2}$&{\bf 14.39}&{\bf 99--~~1}  &{\bf 0.69}&$\boldsymbol{15.072-0.0069\,}$   &{\bf 1}     \\
Clino&$cSiB$ &14.71 &79--~~6 &0.17&$14.933-0.0027\,$    &0.9    \\
\hline
\multicolumn{4}{l}{\bf O--Si--O bend}\\
En--Fs&$\boldsymbol{oOB_1}$&{\bf 17.54}&{\bf 99--~~1} &{\bf 0.81}&$\boldsymbol{18.249-0.0072\,}$     &{\bf 1}     \\
      &$\boldsymbol{oOB_2}$&{\bf 18.28}&{\bf 99--~~1} &{\bf 0.99}&$\boldsymbol{19.285-0.010\,}$     &{\bf 1}     \\
      &$\boldsymbol{oOB_3}$&{\bf 19.42}&{\bf 99--~~1} &{\bf 0.99}&$\boldsymbol{20.393-0.0099\,}$     &{\bf 1}     \\
Clino-&$\mathit{cOB}$  &{\it 19.49}&{\it 79--~~6} &{\it 0.04}&$\mathit{19.520-0.0001\,}$     &{\it 0.2}   \\
En--Wo&$wOB$  &20.66&79--60&-0.04&$20.489+0.0021\,$     &0.9   \\
Dp--Hd&$hOB$  &20.75&50--~~6 &0.48&$21.290-0.011\,$     &1     \\  
\hline
\multicolumn{4}{l}{\bf Translations}\\
En--Fs&$oT_{1E}$ &21.74&99--55&-0.23&$21.240+0.0048\,$    &1     \\
     &$\boldsymbol{oT_2}$ &{\bf 24.57}&{\bf 99--~~1} &{\bf 3.60}&$\boldsymbol{28.130-0.038\,}$&{\bf 1}     \\ 
     &$oT_{3E}$ &26.60&99--85&0.14&$27.769-0.011\,$    &0.8   \\
     &$\boldsymbol{oT_4}$&{\bf 28.25}&{\bf 99--~~1}&{\bf 3.50}&$\boldsymbol{31.766-0.036\,}$&{\bf 1}     \\
     &$\boldsymbol{oT_5}$   &{\bf 35.84}&{\bf 99--~~1} &{\bf 8.02}&$\boldsymbol{44.027-0.082\,}$&{\bf 1}     \\   
     &$\boldsymbol{oT_{6E}}$ &{\bf 43.67}&{\bf 99--55}&{\bf 3.50}& $\boldsymbol{51.45\;\,-0.079\,}$    &{\bf 1}     \\   
     &$\boldsymbol{oT_{6F}}$ &{\bf 55.0}&{\bf 40--~~1} &{\bf 1.23}&$\boldsymbol{56.09\;\,-0.03\,}$     &{\bf 1}     \\
clino-&$\boldsymbol{cT}$   &{\bf 40.32}&{\bf 79--~~6} &{\bf 2.05}&$\boldsymbol{42.25\;\,-0.031\,}$    &{\bf 0.9}   \\   
En--Wo&$wT_1$ &23.75&79--60&0.76&$26.896-0.040\,$    &1     \\
     &$wT_2$ &29.24&79--60&0.35&$30.696-0.018\,$    &1     \\
     &$wT_3$ &31.45&79--47&0.81&$33.213-0.020\,$    &0.9   \\
     &$wT_4$ &44.25&79--47&0.78&$46.26\;\,-0.024\,$    &1     \\
     &$wT_5$ &66.7&79--50&0.44&$67.67\;\,-0.01\,$    &0.8   \\   
Dp--Hd&$hT_1$ &24.27&50--~~6 &0.54&$24.883-0.012\,$    &1     \\
     &$\boldsymbol{hT_2}$ &{\bf 29.50}&{\bf 50--~~6} &{\bf 1.27}&$\boldsymbol{30.912-0.029\,}$    &{\bf 1}     \\
     &$hT_3$ &34.01&50--~~6 &0.24&$34.307-0.0042\,$    &0.7   \\
     &$\boldsymbol{hT_4}$ &{\bf 67.1}&{\bf50--~~6} &{\bf 7.0~~~}&$\boldsymbol{75.20\;\,-0.2\;\; }$    &{\bf 1}     \\ 
\hline
\end{tabular}
\end{minipage}
\end{table*}

\section{Limitations in our data set}
\subsection{Undetectable effect of clinopyroxene on En$_{99}$}
\label{sec:impurities}

\citet{Hofmeister2012} found that the meteoritic sample from which
En$_{99}$ is derived contains 5 per cent of microscopic inclusions
(blebs) of composition Mg$_
{1.10}$Ca$_{0.85}$Na$_{0.02}$Al$_{0.02}$)Si$_2$O$_6
\equiv$Wo$_{43}$En$_{55}$, a level, just above canonical detection
limit in infrared spectra.  Unfortunately we do not have a spectrum of
Wo$_{43}$En$_{55}$ to compare with our data and we do not know if the
thin film made from sample was similarly contaminated. The spectrum of
En$_{99}$ contains extra peaks that do not appear in the rest of the
En--Fs orthopyroxene series. However, this could be because it is of
end-member composition. Bands which look like clinopyroxene peaks are
a shoulder at 14.68$\mu$m ($cSiB$), and fairly prominent peaks at
20.70~$\mu$m ($wOB$ / $hOB$), 34.01~$\mu$m ($hT_3$) and 40.98~$\mu$m
($cT$). These features were also identified by Chihara et al. in their
magnesium-rich synthetic samples which contained no
Wo$_{43}$En$_{55}$, therefore we consider these bands intrinsic to
En$_{99}$ and not the result of contamination.

\subsection{Unobserved bands above 70~$\mu$m}
\label{sec:unobs}
\citet{Bowey2001} detected a weak pair of bands in En$_{85}$ centred
at 71.01~$\mu$m and 73.83~$\mu$m at room temperature which are not
properly distinguishable from the noise in the current data. There are
hints of structure near 73~$\mu$m in En$_{99}$, En$_{92}$, En$_{85}$
and at 75.8~$\mu$m in En$_{55}$. The most likely cause of their
absence is an insufficiently thick film and low detector
signal-to-noise ratios.  It would be particularly useful to
fully-explore this spectral range because \citeauthor{Chihara2002} measured
differences between clino- and orthoenstatite at longer wavelengths,
namely a pairs of bands at 49.2~$\mu$m and 51.6~$\mu$m, and
68.7~$\mu$m and 72.5~$\mu$m in orthoenstatite (Oen; $oEn_{100}$) which
were not observed in clinoenstatite (Cen; cEn$_{100}$); clinoenstatite
had a narrow band at 65.9~$\mu$m. The Cen and Oen spectra were
identical at wavelengths $\la40\mu$m. The longest band they identified
was a peak at 86.1~$\mu$m in En$_0$.

\section{Consistency of room-temperature and low-temperature particulate spectra}\label{sec:compare}
\begin{figure}
\includegraphics*[bb=60 180 450 715, width=\linewidth]{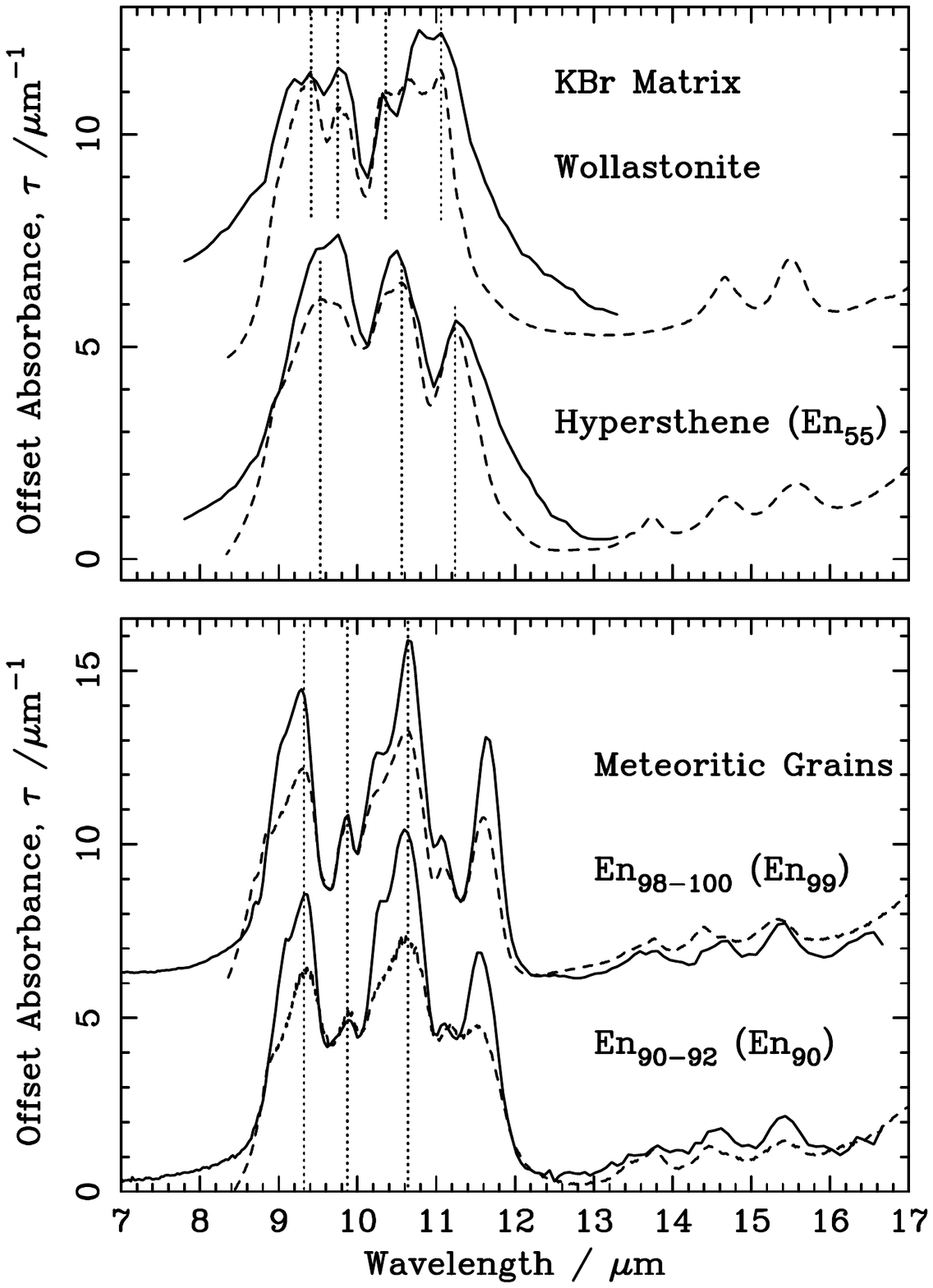}
\caption{Comparison of DAC spectra (dashed curves) with
  \citeauthor{Ferraro1982} KBr-pellet-spectra (solid curves) and
  with those of individual crushed meteoritic grains (solid curves)
  obtained in a Diamond Compression Cell \citep{BMG2007}.  The ratios
  between strong and weak peaks change as does the spectral contrast,
  but the features are not wavelength-shifted provided the minerals
  have similar chemical compositions.\label{fig:rtcomp}}
\end{figure}
\begin{figure}
\includegraphics*[bb=60 395 520 790, width=\linewidth]{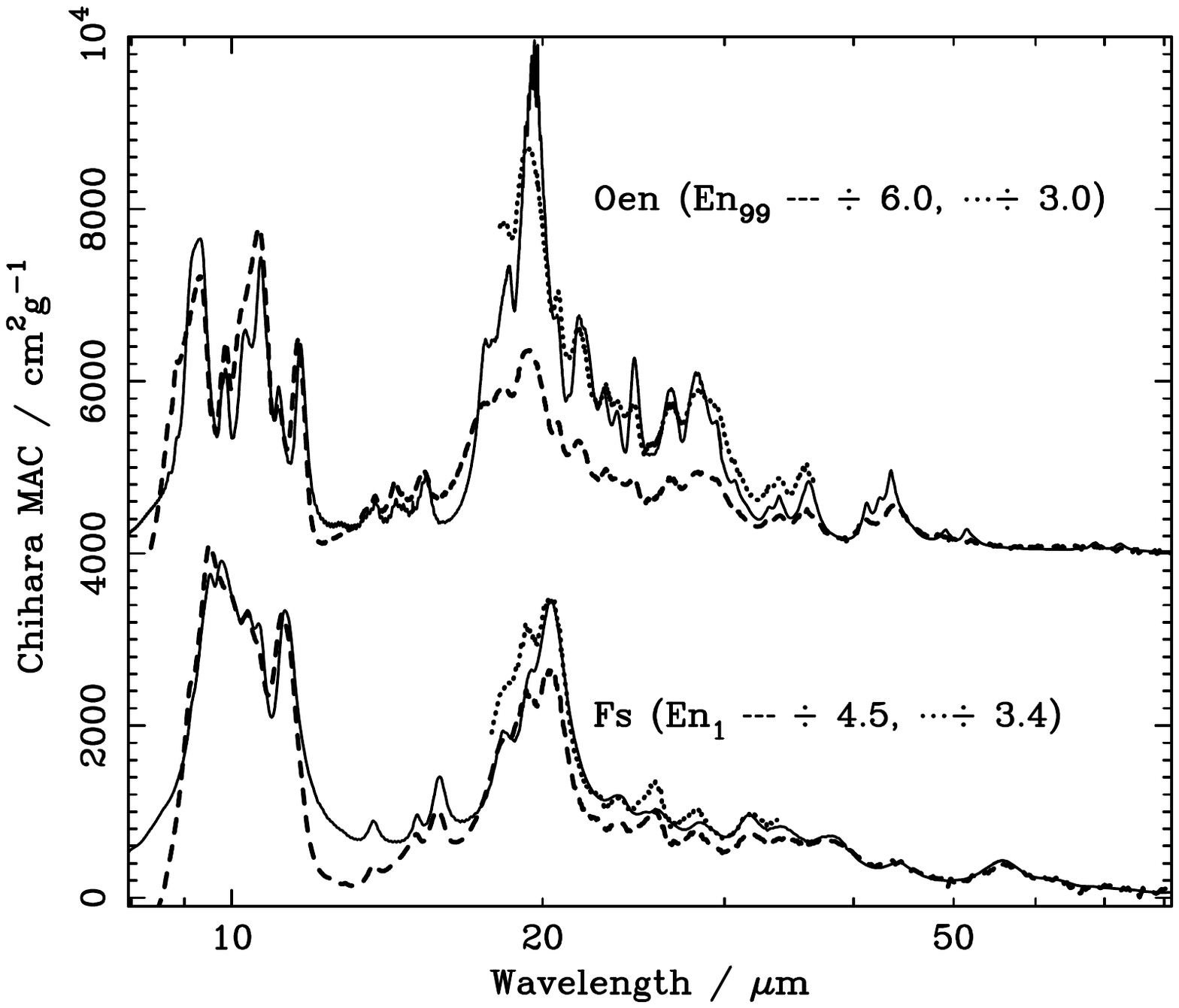}
\caption{Comparison with~\citeauthor{Chihara2002} KBr and polyethylene
  pellet data (solid curves) for synthetic orthoenstatite (Oen$\equiv$
  En$_{100}$) and ferrosilite (Fs$\equiv$ En$_0$). Other curves are
  our data for En$_{99}$ and En$_1$ samples converted to mass
  absorption coefficients scaled to match features in their data.\label{fig:chihara}}
\end{figure}

\subsection{Comparison with room temperature data obtained with KBr and polyethylene pellets}

The DAC spectra are compared with published KBr-pellet spectra of
similar terrestrial pyroxenes in Figures~\ref{fig:rtcomp} and
~\ref{fig:chihara}. The \citeauthor{Ferraro1982} spectra were obtained
with the KBr-pellet technique (grain sizes are unknown, but small
enough to produce consistent peaks) and digitized for comparison with
low-resolution astronomical spectra \citep{BA2002}. The ratios
between strong and weak peaks change as does the spectral contrast,
but features are not wavelength-shifted provided the pyroxenes have
similar chemical compositions. Spectra obtained with the KBr technique
have recognisably similar shapes to the DAC spectra and the peaks are
not wavelength-shifted. The KBr spectra are broader than our data due
to the scattering of light at boundaries between the grains and the
matrix. If differences in refractive index are substantial at grain
boundaries some light will be reflected back along its path as it
passes into the grain and again as it passes back into the KBr. It is
difficult to control particle density and clumping in a KBr
dispersion: if these parameters are too large the measured bandwidths
will increase due to the increased scattering, even if the majority of
grains are sufficiently small. In the DAC the only substantial change
in refractive index occurs at the surface of the diamonds, and its
effect is mitigated by subtracting the spectrum of the empty cell from
the data.

KBr- and polyethylene-pellet spectra of synthetic MgSiO$_3$ (Oen) and
FeSiO$_3$ (Fs) orthopyroxenes obtained by \citeauthor{Chihara2002},
are compared with our En$_{99}$ and En$_{1}$ samples in
Figure~\ref{fig:chihara}. The peak wavelengths quoted by
\citeauthor{Chihara2002} are consistent with our data. However, there
is a substantial difference in our derived mass absorption
coefficients. Their mass absorption coefficients are factors of 6.0
(En$_{99}$) and 4.5 (En$_1$) weaker than ours in the Si--O stretching
and Si--O--Si bending regions (8--17~$\mu$m) and in the far-infrared
beyond $\sim $40~$\mu$m, but the discrepancy in the 17--40~$\mu$m
range for En$_{99}$ is a factor of 3.0 and 3.4. Some large individual
bands are much sharper than ours which is possibly an effect of the
DAC spectra not fully-sampling the c-axis due to the lath-like
pyroxenes lying on their sides when crushed. These discrepancies occur
because infrared spectra depend on the distance photons travel between
grain-scattering and absorption events but the scattering measurement
is complicated by the close correlation of absorption and reflection
when the bands are strong. For this reason, different size grains are
sampled by light transmitted by different motions within the
crystal. DAC and Diamond-Compression-Cell (DCC) measurements (see
Section~\ref{sec:metic}, below) reduce the effect of scattering
because abutted grains have very similar refraction indices so that
the reflection at grain boundaries is minimized and the absence of
scattering enhances the observed mass absorption coefficients.

We also suspect that some of the \citeauthor{Chihara2002} spectra
contain small proportions of quartz (SiO$_2$) in addition to the
olivine they identified because their data show peaks at 12.5~$\mu$m
and 14.4~$\mu$m and enhanced absorption at 9.2~$\mu$m. The effect of
these impurities could be subtracted by the method used in
Appendix~\ref{app:quartz}. None of these issues invalidate their
conclusions and their Oen and Fe spectra contain resolved peaks
longward of 50$\mu$m, and 70~$\mu$m, respectively, when ours do not.

\subsection{Comparison with room temperature spectra of meteoritic grains}
\label{sec:metic}
Meteoritic spectra were obtained by crushing individual micron-sized
grains in a Diamond Compression Cell \citep{BMG2007};
En$_{98-100}$ is normalised average of 13 grains, En$_{90-92}$ is the
average spectrum of 3 grains. These spectra are not wavelength shifted
in comparison with our spectra and the fitted compositional trends are
similar (see Figure~\ref{fig:sioshifts}); where fits are different this
is due to the much smaller compositional range covered by the
meteoritic sample. Spectra of individual meteoritic grains have a much
higher contrast in the strongest Si--O stretches than do the DAC
spectra. Due to this higher contrast and the narrowness of the
features in the tiny samples one or two extra, very weak, bands were
observed in some of the meteoritic spectra.


\subsection{Comparison with Room- and Low-Temperature Spectra}
\begin{figure}
\includegraphics*[bb=60 234 450 715, width=\linewidth]{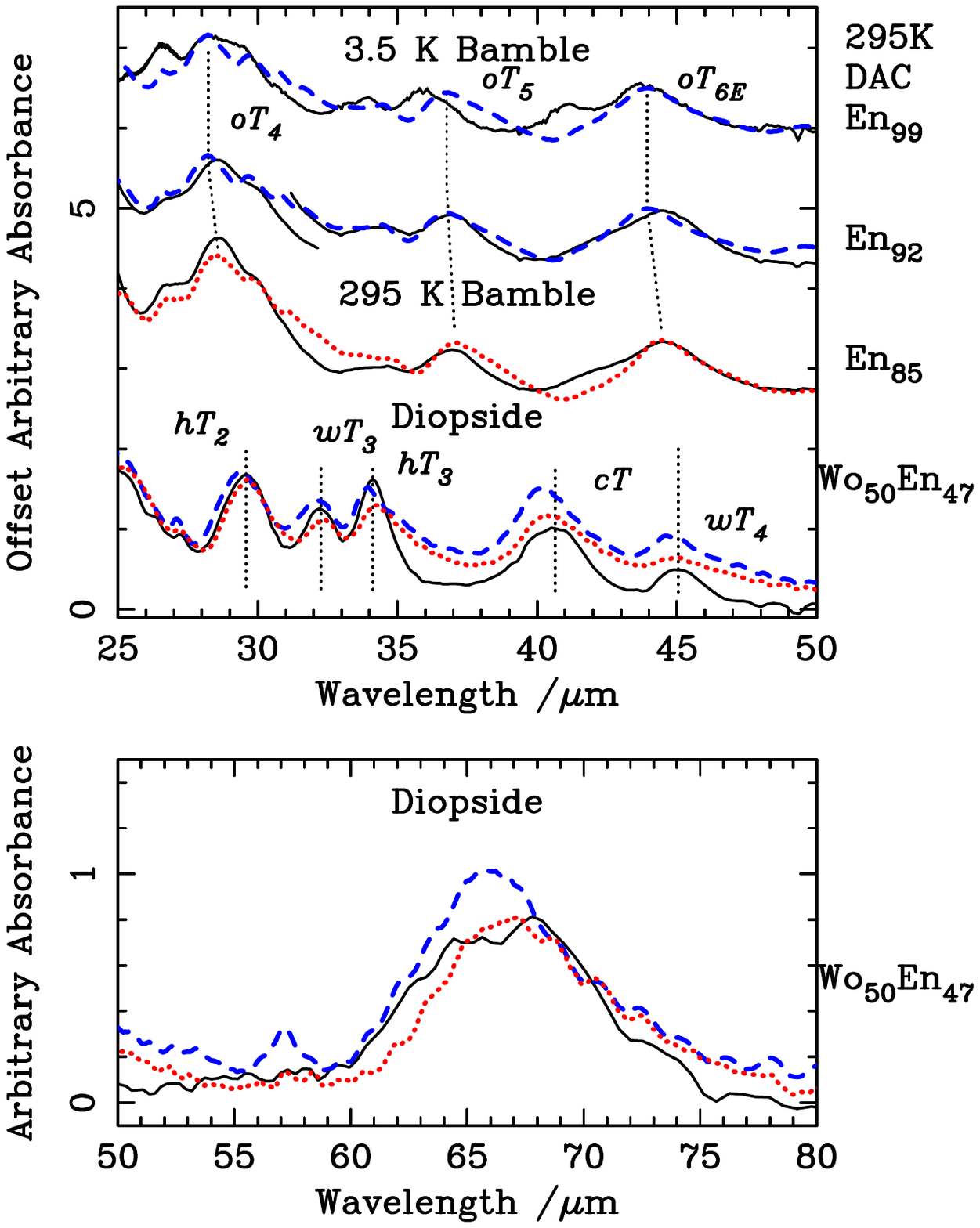}
\caption{Comparison between room temperature DAC Enstatite and
  Wo$_{50}$En$_{47}$ spectra (solid) and \emph{Bamble Enstatite} and
  \emph{Diopside} room temperature (red dotted) and low temperature
  spectra (blue dashed) from \citet{Bowey2001}. The matching DAC
  spectrum is indicated to the right of each spectrum. The difference
  in wavelength between room-temperature spectra of minerals of
  identical composition obtained by the two methods is insignificant
  within the uncertainties. However, the low temperature spectrum of
  En$_{85}$ is blue-shifted with respect to the room temperature
  spectrum making it appear that the cold sample has the composition
  of room temperature En$_{99}$ (bands $oT_{6E}$ and $oT_4$), or
  En$_{92}$ ($oT_5$) , i.e. that it is 7--15\% more Mg-rich than its
  known composition. \label{fig:lowT}}
\end{figure}

Both continuum, and spectral measurements of dust and ice indicate
that the ambient temperature of dust in many astronomical environments
(e.g. planetary nebulae, molecular clouds and the discs of young
stellar objects) is much colder ($\sim 10-200$~K) than the room
temperature spectra obtained in most laboratory experiments. In order
to explore the effect of temperature on the 17-85~$\mu$m spectra of
crystalline olivines and pyroxenes \citet{Bowey2001} obtained
measurements of powdered-samples mixed with petroleum jelly on
0.8-mm-thick polyethylene substrates at room- and low-temperature
($\sim 3.5$~K). They found that 3.5-K peaks occurred at shorter
wavelengths than do the corresponding 295-K peaks.  In order to
compare temperature-related wavelength shifts and compositional
wavelength-shifts their data for pyroxenes are compared with our new
spectra in Figure~\ref{fig:lowT}. As expected, wavelengths of bands and
band shapes of their room-temperature spectra are consistent with our
DAC spectra of samples from the same locality (i.e. samples with near
identical compositions): our spectrum of En$_{85}$ matches their
spectrum of Bamble enstatite, as do Wo$_{50}$En$_{47}$ and the
\citet{Bowey2001} diopside. However, the peak wavelengths of bands
$oT_{6E}$ and $oT_4$ in the 3.5~K spectra are best matched by our
room-temperature En$_{99}$ spectrum and $oT_5$ is best matched by
En$_{92}$ even though the low-temperature sample is En$_{85}$. If the
low-temperature data were from an astronomical source, the conclusion
could be that the enstatite is of near end-member, En$_{100}$,
composition, i.e. it is 7 to 15 per cent richer in Mg than it really
is, or that there are two minerals contributing, En$_{92}$ and
En$_{99}$ when only En$_{85}$ is present.

There is less scope for temperature shifts to masquerade as
compositional shifts in the diopside spectra because bands $hT_2$,
$wT_3$ and $hT_3$ only appear separately in near end-member diopsides
in the Dp--Hd series and the temperature shifts are ten times greater
(0.1--0.3 $\mu$m) than the available compositional shifts and $cT$ is
broader in the non-end-member clinopyroxenes. The longest wavelength
band $hT_4$ is probably too broad and inspecific to pyroxenes to be
useful for temperature studies; differences between the
room-temperature spectra of this mineral are probably due to
uncorrected fringes.

Low-temperature measurements of a wider-variety of pyroxenes are
needed to properly quantify these effects, but the incorrect inferred
compositional differences between our room-temperature and
low-temperature laboratory spectra suggest that Mg-end-member
astronomical enstatites identified in far-infrared astronomical
spectra might not really be of end-member composition. A lower Mg:Fe
ratio in the astronomical dust would be more consistent with the
ratios measured in the meteoritic enstatites sampled by
\citet{BMG2007}.

\section{Summary}
\label{sec:summary}

We present quantitative 8--90-$\mu$m spectra of 8 Mg-- and Fe--bearing
orthopyroxenes, 9 Mg--, Ca-- and Fe--bearing clinopyroxenes and one
Ca--bearing pyroxenoid obtained at room temperature. Spectra were
obtained from thin films of finely-ground minerals crushed in a DAC
without the use of an embedding medium. Pyroxenes have rich infrared
spectra with many narrow peaks. In general, the wavelengths of the
peaks increase as Mg is replaced in the lattice by Ca or Fe. However,
two bands in the En--Fs (orthopyroxene) series shift to shorter
wavelengths as the Fe component increases from 0 percent to
60~per~cent. The most prominent band to do so is at 11.6~$\mu$m when
Mg$\sim 99$~per~cent and 11.2~$\mu$m when Mg$\la40$ per~cent; below
En$_{40}$ the band is more prominent and does not shift further with
an increase in Fe. This 11.2~$\mu$m feature resembles an astronomical
emission or absorption feature which is normally associated with
olivines, forsterite or occasionally PAH absorption.

Only two bands are common to all the pyroxenes studied: an Si--O
stretching mode at 10.22~$\mu$m in En$_{99}$ and an Si--O--Si-bending
mode at 15.34~$\mu$m in En$_{99}$. We suggest that Si--O--Si bending
modes between 13~and 16~$\mu$m hold promise for the identification of
chain silicates in \emph{MIRI} spectra obtained with the \emph{JWST},
because the pyroxene pattern is distinct in this range.

The new spectra are compared with published 8--17~$\mu$m
room-temperature transmission spectra of meteoritic pyroxenes and KBr-
and polyethylene-pellet spectra of terrestrial pyroxenes. The ratios
between strong and weak peaks change as does the spectral contrast,
but the 10-$\mu$m features are not wavelength-shifted provided the
minerals have the same chemical composition and crystal
structure. However, because scattering between grain boundaries is
reduced in DAC measurements of compressed powders, mass absorption
coefficients in our data are up to 6 times the values measured in KBr
and polyethylene pellets.

Room-temperature data from a \citet{Bowey2001} 17--85~$\mu$m study of
two of our samples embedded in petroleum jelly on polyethylene
substrates have band-shapes, wavelengths and relative-band-strengths
which are consistent with our DAC spectra. However, comparison of
their low-temperature spectra with our room-temperature data indicates
that the spectroscopically identified `Mg-end-member' grains in 10~K
astronomical dust might contain 8--15 percent Fe due to
temperature-related wavelength shifts to the blue at low temperatures
masquerading as compositional shifts in peak wavelength. Low- and
intermediate-temperature far-infrared measurements of a wider-variety
of pyroxenes and other silicates are needed to quantify these effects.

Our data contain a wide-variety of spectral features between 40 and
80~$\mu$m which could be used to diagnose both the mineralogy, and the
temperatures of specific grain populations if new observations with a
far-infrared spectrometer on the proposed \emph{Space Infrared
  Telescope for Cosmology and Astrophysics (SPICA)} were to be
obtained. Such a study would provide a substantive link between the
silicates mineralogists and meteoriticists analyse in the laboratory
and the mineralogy inferred from astronomical observations.

\section*{Acknowledgements}
JEB is supported by a 2-year \emph{Science and Technology Research
  Council} Ernest Rutherford Returner Fellowship (ST/S004106/1).  The
measurements at WU by EK and AMH were supported by NSF-AST-9805924 and
NASA APRA04-000-0041. We thank Dan Kremser of Washington University
for the microprobe analyses.

\section*{Data Availability}

The pyroxene spectra presented this article are subject to a partial
embargo of 12 months from the publication date of the article during
which the data will be available from the authors by request. Once the
embargo expires the data will be available from
https://zenodo.org/communities/mineralspectra/







\appendix
\section{Estimated film thicknesses, merging points and quartz correction}

\label{app:merge}

\begin{table}
\caption{Mid-IR (450--4000~\wno spectra were defringed and merged with
  scaled far-IR (80--650~\wno) spectra at point P1. In both cases the
  separation between data-points is about 0.5~\wno, unless otherwise
  specified. The 1--2~\wno-resolution specified in
  Section~\ref{sec:expt} describes how close peaks can lie but still
  be distinguished. In some spectra
  reflection artifically broadened the Si--O stretch near 1100~\wno:
  this was subtracted by fitting a curve to the unaffected spectrum of
  En$_{55}$ and scaling this to the affected spectra at P2; the
  position of P2 was chosen so that real kinks in the data were not
  removed.}
\begin{minipage}{\linewidth}
\begin{tabular}{llll}
\hline
Sample  &$d$ (MIR)&P1&P2\\ 
        &$\mu$m&\wno~($\mu$m)&\wno~($\mu$m)\\
\hline
En$_{99}$&0.25&582~(17.2) &1156~(8.6)\\ 
En$_{92}$&0.29&546~(18.3)&1150~(8.7)\\ 
En$_{90}$&0.38&595~(16.8) &--\\ 
En$_{85}$&0.86&519~(19.3) &--\footnote{Defringing was not required; MIR resolution was 1~\wno; FIR 0.5~\wno}\\ 
Wo$_{10}$En$_{62}$&0.87&484(20.7)&--\\ 
En$_{55}$&0.35&486~(20.6)&--\\ 
En$_{40}$&0.25&584~(17.1)&1116~(9.0)\\ 
En$_{12}$&0.44&471~(21.2)&1103~(9.1)\\ 
En$_{1}$ &0.65&520~(19.2)&1102~(9.1)\\ 
\hline
Wo$_{21}$En$_{79}$&0.78&482~(20.7)&1135~(8.8)\\ 
Wo$_{30}$En$_{70}$&0.53&469~(21.3)&1131~(8.8)\\ 
Wo$_{40}$En$_{60}$&0.48&527~(19.0)\footnote{data between 508 and 528 (18.9--19.7~$\mu$m) were interpolated due to difficulties with the merge; structure here is spurious.}&1131~(8.8)\\ 
\hline
Wo$_{50}$En$_{50}$&0.71&458~(21.8)&1136~(8.8)\\ 
Wo$_{50}$En$_{47}$&0.19&490~(20.4)\footnote{MIR spectral resolution was 1.0~\wno, FIR: 0.5~\wno}&1144~(8.7)\\ 
Wo$_{49}$En$_{36}$&1.01&454~(22.0)&--\\ 
Wo$_{36}$En$_{37}$&0.46&442~(22.6)&--\\ 
Wo$_{47}$En$_{6}$&0.38&557~(18.0)&--\\ 
\hline
Wo$_{99}$En$_1$&0.88   &451~(22.2)&--\\ 
\hline
\end{tabular}
\end{minipage}
\end{table}

\subsection{Quartz correction in En$_{12}$ and En$_1$}
\label{app:quartz}

\begin{figure}
\includegraphics*[bb=60 395 510 775,width=\linewidth]{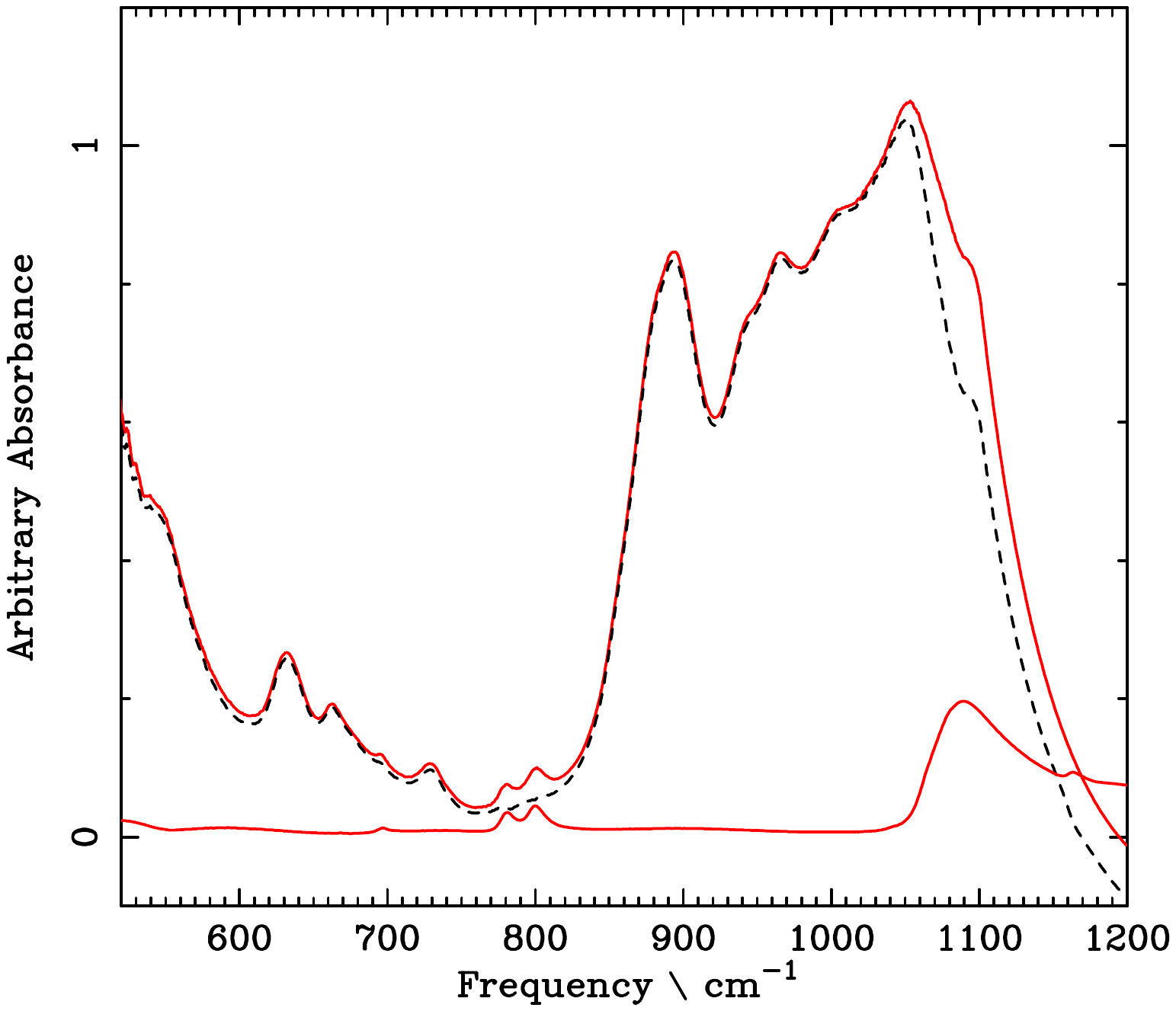}
\caption{Removal of quartz impurity from En$_1$ spectrum. A quartz
  spectrum, scaled to bands at 12.5 and 12.82~$\mu$m (bottom solid
  red curve), was subtracted from the mid-infrared range of En$_1$
  (black dashed line). The cleaned En$_1$ spectrum is the main solid
  red curve.\label{fig:en1corr}}
\end{figure}

The spectrum of En$_{1}$, the ferrosilite, Fe-end-member orthopyroxene
has weak bands at 12.50~$\mu$m, 12.82~$\mu$m (which are very weak in
En$_{12}$) in addition to a very weak 14.37~$\mu$m band. These bands
match the wavelengths of bands in $\alpha$ quartz
\citep[see][]{Koike2013}. Therefore we obtained a DAC spectrum of
quartz, scaled it to the 12.5 and 12.8-$\mu$m features, and subtracted
it from our mid-infrared data (Figure~\ref{fig:en1corr}). These data
were merged with the re-scaled far-infrared spectra before the band
strength calibration. Quartz bands were not detected or subtracted from the
far-infrared spectra.

\begin{table}
\caption{Mineral compositions for samples microprobed at Washington University \label{tab:analysis}}
\begin{minipage}{\linewidth}
\begin{tabular}{llll}
\hline
Sample   &Wo$_{49}$En$_{36}$   & Wo$_{47}$En$_6$        &Wo$_{36}$En$_{37}$\\                            
Locality                        &Calumet &Iona Is.&Belmont\\           
\hline
Oxide / per cent                      &   &                &         \\ 
SiO$_2$                    & 51.57      &48                     &49.95\\
    TiO$_2$                    & 0.16       &-                      &0.85\\
    Al$_2$O$_3$                   & 1.45       &0.5                    &1.83\\
    Cr$_2$O$_3$                   & -          &-                      &0.01\\
    Fe$_2$O$_3$                   & -          &2                      &-     \\
    FeO                     & 7.79       &23                     &14.93 \\
    MnO                     & 0.52       &1                      &0.36  \\
    MgO                     & 13.03      &2                      &13.06 \\
    CaO                     & 24.55      &21                     &17.43 \\
    Na$_2$O                    & 0.11       &-                      &0.25  \\
K$_2$O                     & -          &-                      &-     \\                               
H$_2$O$^+$                    & -          &-                      &-     \\                               
H$_2$O$^-$                    & -          &-                      &-     \\                               
Impurities              &            &                       &      \\                               
\hline
Total / per cent        & 99.18      &97.5                   &98.67 \\                               
\hline

\end{tabular}
\end{minipage}
\end{table}


\bsp	
\label{lastpage}
\end{document}